\input harvmac
\input epsf
\noblackbox

\newcount\figno

\figno=0
\def\fig#1#2#3{
\par\begingroup\parindent=0pt\leftskip=1cm\rightskip=1cm\parindent=0pt
\baselineskip=11pt
\global\advance\figno by 1
\midinsert
\epsfxsize=#3
\centerline{\epsfbox{#2}}
\vskip 12pt
\centerline{{\bf Figure \the\figno} #1}\par
\endinsert\endgroup\par}
\def\figlabel#1{\xdef#1{\the\figno}}
\def\pano{\par\noindent}
\def\smno{\smallskip\noindent}
\def\meno{\medskip\noindent}
\def\bigno{\bigskip\noindent}
\def\ds{\displaystyle}
\font\cmss=cmss10
\font\cmsss=cmss10 at 7pt

\def\rlx{\relax\leavevmode}
\def\inbar{\vrule height1.5ex width.4pt depth0pt}
\def\IC{\relax\,\hbox{$\inbar\kern-.3em{\rm C}$}}
\def\IR{\relax{\rm I\kern-.18em R}}
\def\IN{\relax{\rm I\kern-.18em N}}
\def\IP{\relax{\rm I\kern-.18em P}}
\def\ZZ{\rlx\leavevmode\ifmmode\mathchoice{\hbox{\cmss Z\kern-.4em Z}}
 {\hbox{\cmss Z\kern-.4em Z}}{\lower.9pt\hbox{\cmsss Z\kern-.36em Z}}
 {\lower1.2pt\hbox{\cmsss Z\kern-.36em Z}}\else{\cmss Z\kern-.4em Z}\fi}

\def\narrowplus{\kern -.04truein + \kern -.03truein}
\def\narrowminus{- \kern -.04truein}
\def\narrowminussub{\kern -.02truein - \kern -.01truein}

\def\o#1{\overline{#1}}
\def\la{\langle}
\def\ra{\rangle}



\lref\rcklt{G.~Curio, A.~Klemm, D.~L\"ust, and S.~Theisen, {\it On the Vacuum
     Structure of Type II String Compactifications on Calabi-Yau Spaces with
     H-fluxes}, hep-th/0012213.}

\lref\rapt{I.~Antoniadis, H.~Partouche, and T.~R.~Taylor,
{\it Spontaneous Breaking of N=2 Global Supersymmetry}, 
Phys. Lett. {\bf B372} (1996) 83, hep-th/9512006.}

\lref\rfgp{S.~Ferrara, L.~Girardello, and M.~Porrati,
{\it Minimal Higgs Branch for the Breaking of Half of the Supersymmetries in
N=2 Supergravity}, Phys. Lett. {\bf B366} (1996) 155, hep-th/9510074.} 

\lref\rfgptwo{S.~Ferrara, L.~Girardello, and M.~Porrati,
{\it Spontaneous Breaking of N=2 to N=1 in Rigid and Local Supersymmetric
Theories}, Phys. Lett. {\bf B376} (1996) 275, hep-th/9512180.} 

\lref\rtv{T.~R.~Taylor, and C.~Vafa, {\it RR flux on Calabi-Yau and Partial
Supersymmetry Breaking}, Phys. Lett. {\bf B474} (2000) 130, hep-th/9912152.} 

\lref\rseiwit{N.~Seiberg and  E.~Witten, {\it String Theory and 
Noncommutative Geometry}, JHEP {\bf 9909} (1999) 032, hep-th/9908142.}

\lref\rschomi{V.~Schomerus, {\it D-branes and Deformation Quantization},
JHEP {\bf 9906} (1999) 030, hep-th/9903205.}

\lref\rangles{M.~Berkooz, M.~R.~Douglas and R.~G.~Leigh, {\it Branes Intersecting 
at Angles}, Nucl. Phys. {\bf B480} (1996) 265, hep-th/9606139.}

\lref\rdienes{K. R. Dienes, {\it Solving the Hierarchy Problem without 
Supersymmetry or Extra Dimensions: An Alternative Approach}, 
 hep-ph/0104274.}

\lref\jab{F.~Ardalan, H.~Arfaei and  M.~M.~Sheikh-Jabbari, {\it 
Noncommutative Geometry From Strings and Branes}, 
JHEP {\bf 9902} (1999) 016, hep-th/9810072.}

\lref\rfs{W.~Fischler and L.~Susskind, {\it Dilaton Tadpoles, String
Condensates and Scale Invariance}, Phys. Lett. {\bf B173} (1986) 262.}

\lref\rfstwo{W.~Fischler and L.~Susskind, {\it Dilaton Tadpoles, String
Condensates and Scale Invariance 2}, Phys. Lett. {\bf B171} (1986) 383.}

\lref\rpolgid{S.B. Giddings, S. Kachru and J. Polchinski, {\it
Hierarchies from Fluxes in String Compactifications}, 
hep-th/0105097.}

\lref\rakt{I. Antoniadis, E. Kiritsis and T. Tomaras, {\it
A D-brane alternative to unification}, Phys.Lett. {\bf B486} (2000) 186, 
hep-ph/0004214.}

\lref\rtsey{E.S. Fradkin and A.A. Tseytlin, {\it Effective Field Theory
from Quantized Strings}, Phys.Lett. {\bf B158} (1985) 316;
 A. Abouelsaood, C.G. Callan, C.R. Nappi and S.A. Yost, {\it
Open Strings in Background Gauge Fields},  Nucl.Phys. {\bf B280} (1987) 599.}

\lref\rcgpr{Y.~Chikashige, G.~Gelmini, R.~D.~Peccei, and M.~Roncadelli, {\it
Horizontal Symmetries, Dynamical Symmetry Breaking and Neutrino Masses},
Phys. Lett. {\bf B94} (1980) 499.} 

\lref\wittena{E.~Witten, {\it Search for realistic Kaluza-Klein theory},
Nucl. Phys. {\bf B186} (1981) 412.}

\lref\CHSW{P.~Candelas, G.~Horowitz, A.~Strominger and E.~Witten,
{\it Vacuum configurations for superstrings}, Nucl. Phys. {\bf B258} (1985)
46.} 

\lref\LAT{W.~Lerche, D.~L\"ust and A.~N.~Schellekens, {\it Chiral, 
four-dimensional heterotic strings from self-dual lattices},
Nucl. Phys. {\bf B287} (1987) 477.}

\lref\radd{N.~Arkani-Hamed, S.~Dimopoulos, and G.~Dvali, {\it The Hierarchy
Problem and New Dimensions at a Millimeter}, Phys. Lett. {\bf B429} (1998)
263, hep-ph/9803315.}

\lref\raadd{I.~Antoniadis, N.~Arkani-Hamed, S.~Dimopoulos, and G.~Dvali, {\it
New Dimensions at a Millimeter to a Fermi and Superstrings at a TeV},
Phys. Lett. {\bf B436} (1998) 257, hep-ph/9804398.}

\lref\rrstwo{L.~Randall, and R.~Sundrum, {\it An Alternative to
Compactification}, Phys. Rev. Lett. {\bf 83} (1999) 4690, hep-th/9906064.}

\lref\rrs{L.~Randall, and R.~Sundrum, {\it A Large Mass Hierarchy from a small
Extra Dimension}, Phys. Rev. Lett. {\bf 83} (1999) 3370, hep-th/9905221.}

\lref\FERM{H.~Kawai, D.~Lewellen and S.~H.~Tye, {\it Construction of fermionic
string models in four dimensions}, Nucl. Phys. {\bf B288} (1987) 1;
I.~Antoniadis, C.~Bachas and C.~Kounnas, {\it Four-dimensional superstrings},
Nucl. Phys. {\bf B289} (1987) 87.}

\lref\SING{M.~Douglas and G.~Moore, {\it D-branes, quivers and ALE instantons},
hep-th 9603167;
I.~R.~Klebanov and E.~Witten, {\it Superconformal field theory on three-branes
at a Calabi-Yau singularity}, Nucl. Phys. {\bf B536} (1998) 199,
hep-th/9807080.}

\lref\NS{K.~Landsteiner, E.~Lopez and D.~Lowe, {\it Duality of chiral $N=1$
supersymmetric gauge theories via branes}, JHEP {\bf 9802} (1998) 007,
hep-th/9801002;
I.~Brunner, A.~Hanany, A.~Karch and D.~L\"ust,
{\it Brane dynamics and chiral non-chiral transitions},
Nucl. Phys. {\bf B528} (1998) 197, hep-th/9801017;
A.~Hanany and A.~Zaffaroni, {\it On the realization of chiral four-dimensional
gauge theories using branes}, JHEP {\bf 9805} (1998) 001, 
hep-th/9801134.}

\lref\HORWIT{P.~Horava and E.~Witten, {\it Heterotic and type I string dynamics
from eleven dimensions}, Nucl. Phys. {\bf B460} (1996) 506, 
hep-th/9510209.}

\lref\ORBI{L.~Dixon, J.~A.~Harvey, C.~Vafa and E.~Witten, {\it Strings on
orbifolds}, Nucl. Phys. {\bf B261} (1985) 678.}

\lref\ORBItwo{L.~Dixon, J.~A.~Harvey, C.~Vafa and E.~Witten, {\it Strings on
orbifolds 2}, Nucl. Phys. {\bf B274} (1986) 285.}

\lref\FAUX{P.~Horava and E.~Witten, {\it Heterotic and type I string dynamics
from eleven dimensions}, Nucl. Phys. {\bf B460} (1996) 506, 
hep-th/9510209;
E.~Witten, {\it Strong coupling expansion of  Calabi-Yau compactification},
Nucl. Phys. {\bf B471} (1996) 135, hep-th/9602070;
M.~Faux, D.~L\"ust and B.~A.~Ovrut, {\it Intersecting orbifold
planes and local anomaly cancellation in M-theory},
Nucl. Phys. {\bf B554} (1999) 437, hep-th/9903028;
V.~Kaplunovsky, J.~Sonnenschein, S.~Theisen and S.~Yankielowicz,
{\it On the duality between perturbative orbifolds and M-theory
on $T^4/Z_N$}, hep-th/9912144;
M.~Faux, D.~L\"ust and B.A.~Ovrut, {\it Local Anomaly Cancellation,
M-theory orbifolds and phase transitions}, hep-th/0005251.}

\lref\GAUGINO{J.P.~Derendinger, L.E.~Ibanez and H.P.~Nilles,
{\it On the low-energy $D=4$, $N=1$ supergravity theory extracted from the
$D=10$, $N=1$ superstring}, Phys. Lett. {\bf B155} (1985) 65;
M.~Dine, R.~Rohm, N.~Seiberg and E.~Witten,
{\it Gluino condensation in superstring models},
Phys. Lett. {\bf B156} (1985) 55.}

\lref\rleigh{D. Berenstein, V. Jejjala and  R. G. Leigh, {\it
The Standard Model on a D-brane},  hep-ph/0105042.}

\lref\rBBa{I.~Antoniadis, E.~Dudas and A.~Sagnotti, {\it Brane
supersymmetry breaking}, Phys. Lett. {\bf B464} (1999) 38, 
hep-th/9908023.}

\lref\rBBb{
G.~Aldazabal and A.M.~Uranga, {\it Tachyon-free
Non-supersymmetric Type IIB Orientifolds via Brane-Antibrane Systems}, 
JHEP {\bf 9910} (1999) 024, hep-th/9908072.}

\lref\rBBc{
C.~Angelantonj, I.~Antoniadis, G.~D'Appollonio, E.~Dudas,
A.~Sagnotti, {\it Type I vacua with brane supersymmetry breaking},
hep-th/9911081.}

\lref\rBBd{
C.~Angelantonj, R.~Blumenhagen and M.~R.~Gaberdiel,
{\it Asymmetric orientifolds, brane supersymmetry breaking and non BPS branes},
hep-th/0006033.}

\lref\raiq{G.~Aldazabal, L.~E.~Ibanez and  F.~Quevedo, {\it Standard-like
Models with Broken Supersymmetry from Type I String Vacua}, JHEP 
{\bf 0001} (2000) 031, hep-th/9909172.} 

\lref\raiqtwo{G.~Aldazabal, L.~E.~Ibanez and  F.~Quevedo, {\it A D-Brane
Alternative to the MSSM }, JHEP {\bf 0002} (2000) 015, hep-ph/0001083.} 

\lref\raiqu{G.~Aldazabal, L.~E.~Ibanez, F.~Quevedo and A.~M.~Uranga, 
{\it D-Branes at Singularities: A Bottom-Up Approach to the String
Embedding of the Standard Model}, JHEP {\bf 0008} (2000) 002, hep-th/0005067.} 

\lref\rrab{R.~Rabadan, {\it Branes at Angles, Torons, Stability and
Supersymmetry}, hep-th/0107036.}

\lref\VECTOR{E.~Witten; {\it New issues in manifolds of $SU(3)$ holonomy},
Nucl. Phys. {\bf B268} (1986) 79.}

\lref\WILSON{L.~E.~Ibanez, H.~P.~Nilles and F.~Quevedo,
{\it Reducing the rank of the gauge group in orbifold
compactifications of the heterotic string}, Phys. Lett. {\bf B192} (1987) 332.}

\lref\SM{I.~Antoniadis, E.~Kiritsis and T.N.~Tomaras,
{\it A D-brane alternative to unification},  hep-th/0004214;
A.~Krause, {\it A small cosmological constant, grand unification and warped
geometry}, hep-th/0006226.}

\lref\rfhs{S.~F\"orste, G.~Honecker and R.~Schreyer, {\it Supersymmetric
$\ZZ_N \times \ZZ_M$ Orientifolds in 4-D with D-branes at Angles},
Nucl. Phys. {\bf B593} (2001) 127, hep-th/0008250.}  

\lref\rfhstwo{S.~F\"orste, G.~Honecker and R.~Schreyer, {\it Orientifolds
with Branes at Angles}, JHEP {\bf 0106} (2001) 004, hep-th/0105208.}  

\lref\rbgklnon{R.~Blumenhagen, L.~G\"orlich, B.~K\"ors and D.~L\"ust, 
{\it Noncommutative Compactifications of Type I Strings on Tori with Magnetic
Background Flux}, JHEP {\bf 0010} (2000) 006, hep-th/0007024.}

\lref\rbgklmag{R.~Blumenhagen, L.~G\"orlich, B.~K\"ors and D.~L\"ust, 
{\it Magnetic Flux in Toroidal Type I Compactification}, Fortsch. Phys. 49
(2001) 591, hep-th/0010198.}

\lref\rbkl{R.~Blumenhagen, B.~K\"ors and D.~L\"ust, 
{\it Type I Strings with $F$ and $B$-Flux}, JHEP {\bf 0102} (2001) 030,
hep-th/0012156.}  

\lref\rckkl{G.~Curio, A.~Klemm, B.~K\"ors and D.~L\"ust, 
{\it Fluxes in Heterotic and Type II Compactifications}, hep-th/0106155.} 

\lref\rbgkl{R.~Blumenhagen, L.~G\"orlich, B.~K\"ors and D.~L\"ust, 
{\it Asymmetric Orbifolds, Noncommutative Geometry and Type I
Vacua}, hep-th/0003024.}

\lref\rbgka{R.~Blumenhagen, L.~G\"orlich and B.~K\"ors, 
{\it Supersymmetric Orientifolds in 6D with D-Branes at Angles},
Nucl. Phys. {\bf B569} (2000) 209, hep-th/9908130.}

\lref\rbgkb{R.~Blumenhagen, L.~G\"orlich and B.~K\"ors, 
{\it A New Class of Supersymmetric Orientifolds with D-Branes at
Angles}, hep-th/0002146.}

\lref\rbgkc{R.~Blumenhagen, L.~G\"orlich and B.~K\"ors, {\it
Supersymmetric 4D Orientifolds of Type IIA with D6-branes at Angles},  
JHEP {\bf 0001} (2000) 040, hep-th/9912204.}

\lref\rprada{G.~Pradisi, {\it Type I Vacua from Diagonal
$\ZZ_3$-Orbifolds}, Nucl. Phys. {\bf B575} (2000) 134.}

\lref\rpradb{G.~Pradisi, {\it Type-I Vacua from Non-geometric Orbifolds},
hep-th/0101085.} 

\lref\rbimp{M.~Bianchi, J.~F.~Morales and G.~Pradisi, {\it
Discrete Torsion in Non-geometric Orbifolds and their Open-string
Descendants}, Nucl. Phys. {\bf B573} (2000) 314, hep-th/9910228.}

\lref\rba{C.~Angelantonj, R.~Blumenhagen, {\it Discrete Deformations in 
Type I Vacua}, Phys. Lett. {\bf B473} (2000) 86, 
hep-th/9911190.} 

\lref\rchu{C.-S.~Chu, {\it Noncommutative Open String: Neutral and Charged},
hep-th/0001144.}

\lref\rchen{B.~Chen, H.~Itoyama, T.~Matsuo and K.~Murakami, {\it
p-p' System with B-field, Branes at Angles and Noncommutative Geometry}, 
{\tt hep-th/9910263}.}

\lref\raadds{C.~Angelantonj, I.~Antoniadis, G.~D'Appollonio, E.~Dudas,
A.~Sagnotti, {\it Type I vacua with brane supersymmetry breaking},
hep-th/9911081, to appear in Nucl. Phys. {\bf B}.}

\lref\ras{C.~Angelantonj, A.~Sagnotti, {\it Type I 
Vacua and Brane Transmutation}, hep-th/0010279.}

\lref\raads{C.~Angelantonj, I.~Antoniadis, E.~Dudas, A.~Sagnotti, {\it Type I
Strings on Magnetized Orbifolds and Brane Transmutation},
Phys. Lett. {\bf B489} (2000) 223, hep-th/0007090.}

\lref\rads{I.~Antoniadis, E.~Dudas, A.~Sagnotti, {\it Brane
supersymmetry breaking}, Phys. Lett. {\bf B464} (1999) 38, 
hep-th/9908023.}

\lref\rbfl{C.~Angelantonj, {\it Non-tachyonic open descendants of the
0B string theory}, Phys. Lett. {\bf B444} (1998) 309, 
hep-th/9810214; R.~Blumenhagen, A.~Font and D.~L\"ust, {\it Tachyon free
orientifolds of type 0B strings in various dimensions},
Nucl. Phys. {\bf B558} (1999) 159, hep-th/9904069;
R.~Blumenhagen, A.~Font, A.~Kumar and D.~L\"ust, {\it Aspects of type 0
string theory}, Class. Quant. Grav. {\bf 17} (2000) 989, hep-th/9908155;
R.~Blumenhagen and A.~Kumar, {\it A Note on Orientifolds and Dualities of Type
0B String Theory}, Phys.Lett. {\bf B464} (1999) 46, hep-th/9906234.}

\lref\rss{J.~Scherk, J.H.~Schwarz, {\it Spontaneous breaking of
supersymmetry through dimensional reduction}, Phys. Lett. {\bf B82}
(1979) 60; {\it How to get masses from extra dimensions}, 
Nucl. Phys. {\bf B153} (1979) 61;
E.~Cremmer, J.~Scherk, J.H.~Schwarz, {\it Spontaneously
broken ${\cal N}=8$ supergravity}, Phys. Lett. {\bf B84} (1979) 83;
S.~Ferrara, C.~Kounnas, M.~Porrati, F.~Zwirner, {\it Superstrings
with spontaneously broken supersymmetry and their effective theories},
Nucl. Phys. {\bf B318} (1989) 75.} 

\lref\rssst{R.~Rohm, {\it Spontaneous supersymmetry breaking in
supersymmetric string theories}, Nucl. Phys. {\bf B237} (1984) 553; 
C.~Kounnas, M.~Porrati, {\it Spontaneous supersymmetry breaking in
string theory}, Nucl. Phys. {\bf B310} (1988) 355; 
S.~Ferrara, C.~Kounnas, M.~Porrati, F.~Zwirner, {\it Superstrings
with spontaneously broken supersymmetry and their effective theories},
Nucl. Phys. {\bf B318} (1989) 75; 
C.~Kounnas, B.~Rostand, {\it Coordinate dependent compactifications
and discrete symmetries}, Nucl. Phys. {\bf B341} (1990) 641;
I.~Antoniadis, C.~Kounnas, {\it Superstring phase transition at
high temperature}, Phys. Lett. {\bf B261} (1991) 369;
E.~Kiritsis, C.~Kounnas, {\it Perturbative and non-perturbative
partial supersymmetry breaking: ${\cal N} = 4 \to {\cal N} = 2 \to
{\cal N} =1$}, Nucl. Phys. {\bf B503} (1997) 117, 
hep-th/9703059.} 

\lref\wise{J.~Gomis, T.~Mehen and M.B.~Wise,
{\it Quantum field theories with compact noncommutative extra dimensions},
hep-th/0006160.}

\lref\rbbh{R.~Blumenhagen, V.~Braun, and R.~Helling, {\it Bound States of
D$(2p)$-D0 Systems and Supersymmetric $p$ Cycles}, Phys. Lett. {\bf B510}
(2001) 311, hep-th/0012157.} 

\lref\rbfa{R.~Blumenhagen, A.~Font, {\it Dilaton Tadpoles, Warped Geometries
and Large Extra Dimensions for Nonsupersymmetric Strings}, Nucl. Phys. {\bf
B599} (2001) 241, hep-th/0011269.}

\lref\rdma{E.~Dudas, J.~Mourad, {\it Brane Solutions in Strings with Broken
Supersymmetry and Dilaton Tadpoles}, Phys. Lett. {\bf B486} (2000) 172,
hep-th/0004165.}

\lref\rsild{M. Dine and  E. Silverstein, {\it
 New M-theory Backgrounds with Frozen Moduli},  hep-th/9712166.}

\lref\rssop{I.~Antoniadis, E.~Dudas, A.~Sagnotti, 
{\it Supersymmetry breaking, open strings and M-theory},
Nucl. Phys. {\bf B544} (1999) 469, hep-th/9807011; 
I.~Antoniadis, G.~D'Appollo\-nio, E.~Dudas, A.~Sagnotti, {\it Partial
breaking of supersymmetry, open strings and M-theory},
Nucl. Phys. {\bf B553} (1999) 133, hep-th/9812118; 
{\it Open descendants of $\ZZ_2 \times \ZZ_2$ freely acting
orbifolds}, Nucl. Phys. {\bf B565} (2000) 123, hep-th/9907184.} 

\lref\rmb{A.~Sagnotti, {\it Anomaly cancellations and open-string
theories}, hep-th/9302099; 
G.~Zwart, {\it Four-dimensional ${\cal N}=1$ $\ZZ_N \times \ZZ_M$
orientifolds}, Nucl. Phys. {\bf B526} (1998) 378, 
hep-th/9708040; 
Z.~Kakushadze, G.~Shiu, S.-H.H.~Tye, {\it Type IIB orientifolds,
F-theory, type I strings on orbifolds and type I - heterotic duality},
Nucl. Phys. {\bf B533} (1998) 25, {\tt hep-th/9804092};
G.~Aldazabal, A.~Font, L.~E.~Ibanez, G.~Violero, {\it $D=4$, ${\cal
N} =1$ type IIB orientifolds}, Nucl. Phys. {\bf B536} (1998) 29,
{\tt hep-th/9804026}.}

\lref\rafiruph{G.~Aldazabal, S.~Franco, L.~E.~Ibanez, R.~Rabadan, A.~M.~Uranga,
{\it Intersecting Brane Worlds}, JHEP {\bf 0102} (2001) 047, hep-ph/0011132.} 

\lref\rafiru{G.~Aldazabal, S.~Franco, L.~E.~Ibanez, R.~Rabadan, A.~M.~Uranga,
{\it $D=4$ Chiral String Compactifications from Intersecting Branes},
hep-th/0011073.}  

\lref\rimr{L.~E.~Ibanez, F.~Marchesano, R.~Rabadan, {\it Getting just the
Standard Model at Intersecting Branes}, hep-th/0105155.} 

\lref\rau{G.~Aldazabal, A.~M.~Uranga, {\it Tachyon-free
non-supersymmetric type IIB orientifolds via brane-antibrane systems}, 
JHEP {\bf 9910} (1999) 024, hep-th/9908072.}

\lref\rbac{C.~Bachas, {\it A Way to Break Supersymmetry},
hep-th/9503030.}
 
\lref\rbist{M.~Bianchi, Ya.~S.~Stanev, {\it Open strings on the Neveu-Schwarz
penta-brane}, Nucl. Phys. {\bf B523} (1998) 193, hep-th/9711069.}

\lref\rabpss{C.~Angelantonj, M.~Bianchi, G.~Pradisi, A.~Sagnotti and
Y.~Stanev, {\it Chiral asymmetry in four-dimensional open string vacua},
Phys. Lett. {\bf B385} (1996) 96, hep-th/9606169.}

\lref\rang{C.~Angelantonj, {\it Comments on Open String Orbifolds with a
Non-vanishing $B_{ab}$}, Nucl. Phys. {\bf B566} (2000) 126, hep-th/9908064.}  

\lref\rbps{M.~Bianchi, G.~Pradisi, and A.~Sagnotti, {\it Toroidal
Compactification and Symmetry Breaking in Open String Theories},
Nucl. Phys. {\bf B376} (1992) 365.}

\lref\rkaku{Z. Kakushadze, G. Shiu and  S.-H.H. Tye, {\it 
Type IIB Orientifolds with NS-NS Antisymmetric Tensor Backgrounds},
Phys.Rev. {\bf D58} (1998) 086001, hep-th/9803141.}

\Title{\vbox{
 \hbox{HU--EP--01/28}
 \hbox{NSF-ITP-01-73}
 \hbox{hep-th/0107138}}}
{\vbox{\centerline{The Standard Model from Stable} 
\vskip 0.4cm
      {\centerline{Intersecting Brane World Orbifolds} }
}}
\centerline{Ralph Blumenhagen, 
Boris K\"ors, 
Dieter L\"ust, 
and Tassilo Ott\footnote{$^1$}{{\tt e-mail:
blumenha, koers, luest, ott@physik.hu-berlin.de}}
} 
\bigskip
\centerline{\it Humboldt-Universit\"at zu Berlin, Institut f\"ur  
Physik,}
\centerline{\it Invalidenstrasse 110, 10115 Berlin, Germany}
\smallskip
\bigskip
\centerline{\bf Abstract}
\noindent
We  analyze the perturbative 
stability of non-supersymmetric intersecting 
brane world models on tori. Besides the dilaton tadpole, a dynamical 
instability in the complex 
structure moduli space occurs at string disc level, which drives the
background geometry to a degenerate limit. 
We show that in  certain orbifold models this latter instability is absent
as the relevant moduli are frozen.  
We construct explicit examples of such orbifold intersecting brane
world models and discuss the phenomenological implications
of a three generation Standard Model which descends naturally from an 
$SU(5)$ GUT theory. 
It turns out that various phenomenological issues require the 
string scale to be at least of  the order of the GUT scale. 
As a major difference compared to the Standard Model, 
some of the  Yukawa couplings are excluded
so that the  standard electroweak Higgs mechanism
with a fundamental Higgs scalar is not  realized in this set-up.

\bigskip

\Date{07/2001}
\newsec{Introduction}
\smno
During the last years string theory has provided a lot of new insights
into fundamental issues of theoretical physics, such as the relation
between gauge theories and gravity or
geometry, the quantum nature of black holes and
the appearance of non-commutative space-time structures. 
Also concerning more phenomenological questions strings proved themselves to
be rather fruitful, perhaps most notably in the context of string
compactifications with large extra dimensions \refs{\radd,\raadd} or localized 
gravity on a four-dimensional domain wall \refs{\rrs,\rrstwo}. D-branes and
non-perturbative duality symmetries always played a key role in all these
developments. Nevertheless, still it is a great challenge to derive the
observed physics of the Standard Model of particle physics directly from
strings. 

Recently a class of string compactifications was investigated which comes
relatively close to the goal of obtaining just the Standard Model 
from strings.
These models are given by type I string compactifications 
on a six-dimensional torus $T^6$ with D9-branes where internal background gauge
fluxes on the branes are turned on
\refs{\rbgklnon\raads\rbgklmag\ras\rafiru\rafiruph\rbkl-\rimr}.\footnote{$^1$}{For
alternative compactifications with D-branes in type I or type II string
theory see 
\refs{\raiq\raiqtwo\rakt\raiqu-\rleigh}; 
a recent discussion
of heterotic string compactifications with background gauge fluxes and
their relation to type II compactifications with internal H-fluxes can be found
in \rckkl. Non-supersymmetric string models with background RR fluxes
were discussed in \rpolgid.} 
Thus, at tree level supersymmetry is only broken on the D-branes with
the bulk still preserving some supersymmetry \refs{\rBBa\rBBb\rBBc-\rBBd}.
Turning on magnetic flux  has the effect that the
coordinates of the internal torus become non-commutative.
In a T-dual picture one is dealing with D6-branes which wrap 3-cycles
of the dual torus and intersect each other at certain angles, determined
by the original gauge fluxes. In this way it was possible to construct string
models with three generations of quarks and leptons and Standard Model
gauge group $SU(3)\times SU(2)_L\times
U(1)_Y$, where supersymmetry is broken on the branes by the gauge
fluxes or, in the dual picture, by the different intersection angles.

Let us recall in slightly more detail the main features of these type I 
string models. Following the old ideas of \rtsey\ and \rbac\ it was first 
described in \rbgklnon\ in a pure stringy language how 
type I compactifications with background fluxes or intersecting branes 
lead to a 
reduction of the gauge group, to chiral fermions and to broken supersymmetry on
the branes. A nice geometrical feature of such models is that the number
of chiral fermions which are localized
at the intersection points of the D6-branes is simply determined  by the 
corresponding topological intersection number of the branes.
In this way a model with four generations of quarks and leptons and
a Standard Model gauge group was obtained in \rbgklnon. 
Later it was shown in \rbkl\ how
odd numbers of generations arise, in particular three, if one adds
to the gauge fluxes also a quantized background NSNS $B$-field
\refs{\rbps,\rkaku,\rang,\ras}.  
In the T-dual picture the torus is then no longer rectangular but tilted
by a discrete angle. In \refs{\rafiru,\rafiruph} 
additional type II models with backgrounds of the form $T^{2d}\times 
(T^{6-2d}/\ZZ_N)$ were considered where the 
D$(3+d)$-branes wrap only $d$-cycles
of the first torus and are point-like on the orbifold. In this way it is
possible that the orbifold space, which is transversal to the branes,
becomes large, whereas the large extra dimension scenario is in conflict
with chirality for the case of $T^6$ compactifications \rbgklnon.
Finally, in \rimr\ a systematic analysis was provided how to obtain
$T^6$ models with precisely three generations of quarks and leptons and
just the Standard Model gauge group without any extension. 
In this context the mass generation
for $U(1)$ gauge bosons due to  flux-induced Green-Schwarz terms and
the related issue of chiral $U(1)$ anomalies is very important.
Furthermore the question how to avoid open string
tachyons in a certain range of the toroidal background parameters and other
phenomenological issues were also addressed.

For type I compactifications the main consistency restrictions
considered so far come from the requirement of the absence of massless tadpoles
in the Ramond-Ramond (RR) sector of the theory. Having no RR-tadpoles
ensures the anomaly freedom in the effective field theory of the massless
modes. As mentioned already, 
the absence of open string tachyons is another important constraint
for model building, where however some `tachyons' might be even welcome
from the phenomenological point of view, namely those which contribute
to the required spontaneous gauge symmetry breaking, in particular
of $SU(2)_L\times U(1)_Y\rightarrow U(1)_{\rm em}$. Hence, one may want to
identify the standard Higgs field with a tachyon, and also those of other
spontaneously broken local gauge groups, such as for instance  $U(1)_{B-L}$. 

In this paper we like to emphasize
that all models considered so far are generically 
unstable due to the existence
of NSNS closed string tadpoles. Specifically we will see that, already
at the topology of the world sheet disc amplitude, the NSNS tadpoles
related to the closed string moduli $U^I$,
the complex structure deformations of $T^6$, and those related to
the closed string dilaton $\phi$ are non-vanishing.
This means that in the induced effective potential these scalar fields
do not acquire a stable minimum but show the typical runaway behaviour.
The explicit form of the potential implies that  the internal geometry 
is driven to a  degenerate singular limit,  where 
all D-branes finally lie on top of each other. As a
result, space-time supersymmetry is  reenforced. In addition,
the dilaton tadpole drives the theory to weak coupling.
By T-duality this also disproves the existence of partial supersymmetry
breaking in type I vacua after introducing magnetic fluxes into the toroidal
${\cal N}=4$ compactification. Similar partial breaking from ${\cal N}=2$ to
${\cal N}=1$  has been shown to be possible in
heterotic and type II theories, albeit under very special circumstances only
\refs{\rapt\rfgptwo\rtv-\rcklt}.  

There are essentially two string theoretic methods  to cure the problem 
of the NSNS tadpoles at least at the next to leading order.
First one can employ  the Fischler-Susskind mechanism \refs{\rfs,\rfstwo}, 
by which 
the back-reaction of the massless fields on the NSNS tadpoles 
is taken into account iteratively. As demonstrated in \refs{\rdma,\rbfa}
solving the string 
equations of motions including the one-loop dilaton tadpole 
in general leads to warped geometries and non-trivial 
profiles of the dilaton and other scalar fields. 
Moreover, in the non-supersymmetric type I string theory discussed in   
\refs{\rdma,\rbfa} the phenomenon of 
spontaneous compactification occurred due to the
NSNS tadpoles. Of course, at this state one is stuck again, as
technically the non-linear sigma model in this highly curved backgrounds
can not be solved exactly. If it could be solved, one would certainly detect
a non-vanishing tadpole at the next order in the string coupling constant. 
Thus, one might hope that the non-supersymmetric string theory self-adjust
its background  order by order in string perturbation theory 
until eventually the true quantum vacuum with vanishing tadpoles to all
orders is reached \rdienes.

A second less ambitious approach to handle at least some of the 
tadpoles is simply by freezing the dangerous closed string 
scalar fields to fixed values. This can be achieved by performing
appropriate projections in an orbifold theory.
In the following we will focus on this second approach. 
In particular we will construct non-supersymmetric
orbifold intersecting brane models, where the 
complex structure moduli of the torus are fixed, and hence
there are no associated NSNS tadpoles. However, the dilaton tadpole
will still survive, and it cannot be excluded that 
new tadpoles will  be induced at higher orders in string perturbation theory. 
As noted in \rsild,  in the M-theory context one can even contemplate
on orbifold actions which freeze the size of the eleventh direction
and therefore of the dilaton in the dual string theory. 
We will further outline the strategy how to obtain orbifold models
with three generations of quarks and leptons and with Standard Model
gauge group $SU(3)\times SU(2)_L\times U(1)_Y$ in this particular kind of
background.  

Our work will be organized as follows. In the next section we will review
the main ingredients of the toroidal intersecting brane worlds.
Next, in section three, we will extract for the toroidal case
all NSNS tadpoles from the infrared divergences in the tree channel
Klein-bottle, annulus and M\"obius strip amplitudes, and we will compute
the corresponding scalar potential.
In section four we will construct $\ZZ_3$ orbifold intersecting brane models
which are free of geometric NSNS tadpoles, especially addressing the 
form of the massless spectrum and the question of anomaly cancellation.
These results  will  be analyzed  in chapter five to find models 
which come as close
as possible to the Standard Model with three generations. 

Unlike the
previous toroidal constructions, where the Standard Model fermions originate
from bifundamental open strings states, we will now be forced to
realize the right-handed $(u,c,t)$-quarks in the antisymmetric representation
of $U(3)$. 
With this assignment it is indeed possible to
get models with three Standard Model generations. 
We will also discuss the open string tachyons of the theory to see whether
the Standard Model Higgs and another Higgs breaking $U(1)_{B-L}$ 
can be realized as tachyons. 
It turns out that all models with an appropriate Higgs scalar descend from a
GUT theory where  $SU(3)\times SU(2)_L\times U(1)_Y \times U(1)_{B-L}$ can be 
unified into $SU(5)\times U(1)$ by a deformation which is marginal at tree
level. The additional global symmetries prohibit the usual Yukawa couplings of
the $(u,c,t)$ quarks and the Standard Model Higgs doublets, so that the
standard mass generation mechanism with fundamental scalar Higgs 
fields does not work. Furthermore, we analyze 
the unification behaviour of gauge couplings and the possibility of proton
decay in the context of the $SU(5)\times U(1)$ GUT model.
In an appendix we also include the results for similar six-dimensional models
providing an extra consistency check for our formalism 
via the six-dimensional anomaly cancellation conditions. 

\newsec{Intersecting brane worlds}
\smno
In this section we review the construction of generically 
non-supersymmetric open string vacua with D-branes intersecting at angles. 
Our starting point is
an ordinary type I model, where for simplicity we consider
only toroidal orbifold models, which we can write as
\eqn\typeone{  {{\rm Type\ IIB}\ {\rm on}\ T^{2d} \over \{G+\Omega G\} }
,} 
where $G$ is a finite group acting on the $2d$ dimensional torus $T^{2d}$. 
In the following we restrict ourselves to the case that the 
closed string sector of the orientifold model \typeone\ 
preserves some supersymmetry. 
Usually, tadpole cancellation requires the introduction of D-branes,
which can be chosen to be BPS so that the open string sector
preserves the same supersymmetry as the closed string sector. 
However, RR tadpole cancellation alone does not require
the open string sector to be  supersymmetric. As shown in
\refs{\rbgklnon,\rbgklmag} for the toroidal case ($G=1$),
there exists the possibility of turning on various constant magnetic 
$U(1)$ fluxes on the
D-branes without  giving up RR tadpole cancellation.
This breaks supersymmetry, and one
faces the usual problems with non-supersymmetric string theories
like tachyons, dilaton tadpoles, moduli stabilization and
the cosmological constant problem.  

Technically, it turned out to be more appropriate to describe such models
in a T-dual language, where the new degrees of freedom are 
described in a purely geometric manner. 
Let us assume that the $2d$-dimensional torus can be written
as a product of $d$ two-dimensional tori 
\eqn\split{ T^{2d}=\bigotimes_{I=1}^d   T_I^{2} ,}
where on each $T_I^2$ we introduce a complex coordinate $Z_I=X_I+i Y_I$. 
Applying T-duality $T_Y$ 
to the $d$ $Y_I$-directions of the $d$ two-dimensional tori,
the orientifold model \typeone\ is mapped to \rbgkl 
\eqn\typeoneb{  {{\rm Type\ II}\ {\rm on}\ T^{2d} \over 
\{\hat{G}+\Omega {\cal R}\hat{G}\} } ,}
where ${\cal R}$ is the reflection of the $Y_I$ and $\hat{G}$ is the image
of $G$ under T-duality $\hat{G}=T_Y\,G\, T_Y^{-1}$.
For the case that $G=\ZZ_N$ the symmetry group acts on each torus 
by rotations 
\eqn\phase{ Z^L_I\to e^{2\pi i {v_I/ N}}\,  Z^L_I, \quad\quad
            Z^R_I\to e^{2\pi i {v_I/N}}\,  Z^R_I .}
Supersymmetric models have been classified in terms of $v_I$ in
\refs{\ORBI,\ORBItwo}.  
The T-dual action is then given by 
\eqn\phaset{ Z^L_I\to e^{2\pi i {v_I/N}}\,  Z^L_I, \quad\quad
            Z^R_I\to e^{-2\pi i {v_I/N}}\,  Z^R_I .}
Thus, T-duality exchanges left-right symmetric actions with
left-right asymmetric actions. 
Moreover, under T-duality D9$_a$-branes with constant 
magnetic fluxes $F^I_a$ are
mapped to D$(9-d)$-branes intersecting at relative angles \rangles
\eqn\angles{
\varphi^I_{ab} = {\rm arctan} (F^I_a) - {\rm arctan} (F^I_b) .}  
They are wrapped around 
one-dimensional cycles on each $T_I^2$, so that each brane $a$ is specified
by two coprime wrapping  numbers $(n_a^I,m_a^I)$ for each torus. 
Moreover, the $\Omega{\cal R}$ symmetry allows two inequivalent
choices of the complex structure 
\eqn\compl{    U^I=U_1^I-i\, U_2^I={{e_1} \over {e_2} } 
=U_1^I-i\,{R_1^I\over R_2^I} }
of each $T^2$, $U^I_1=0$ or $1/2$. The tori are 
depicted in figure 1.
\fig{}{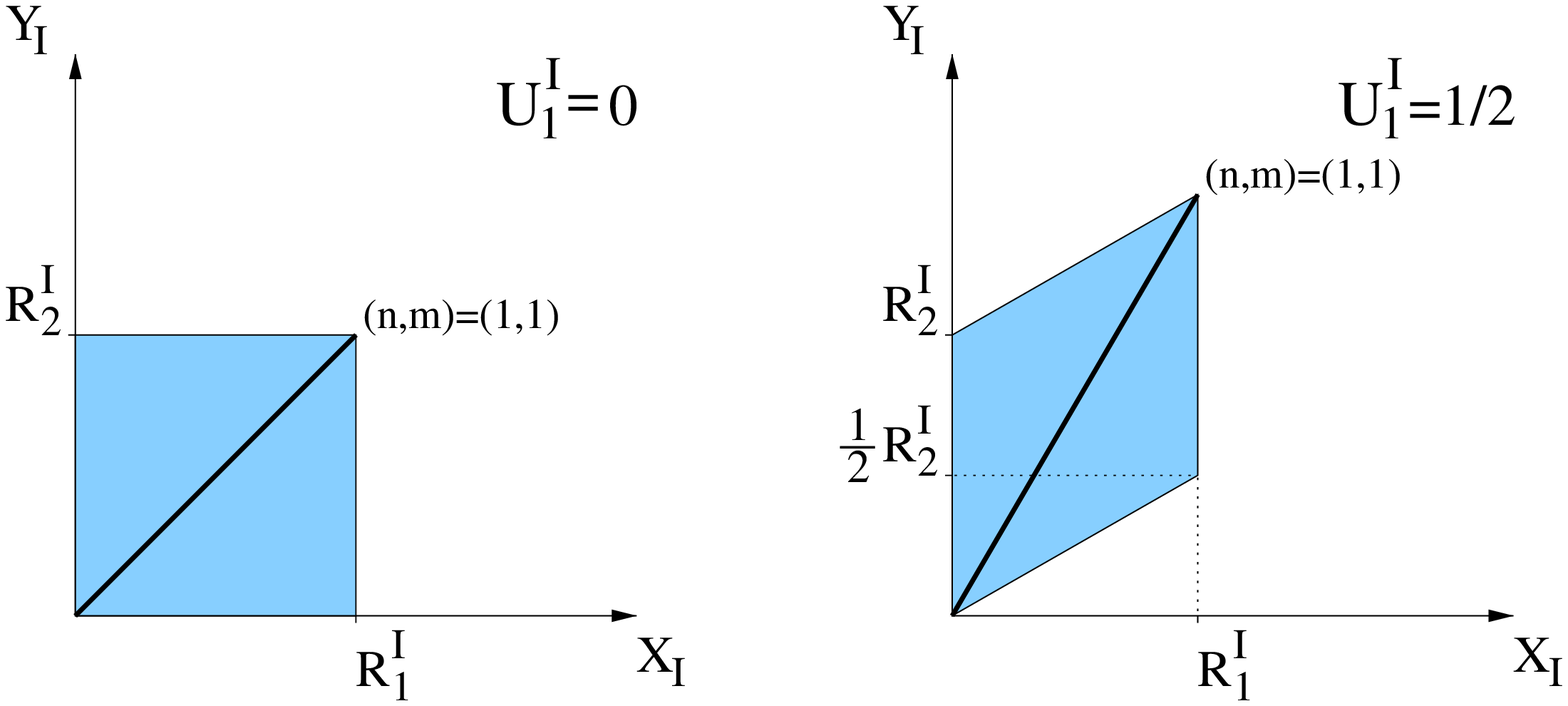}{16truecm}
In the T-dual picture with magnetic fluxes the tilt of the torus
corresponds to turning on a discrete NSNS two-form $B$-field. 
One also has to take into account that for each brane D$_a$ there
must exist the mirror brane D$_{a'}$, which is its image under 
$\Omega{\cal R}$. For the purely toroidal case, 
often called type I$'$, the RR-tadpole cancellation conditions 
were derived in \rbgklnon. 
If one introduces $K$ stacks of D6$_a$-branes counted together with their
$\Omega{\cal R}$ mirrors, then  the four dimensional   RR-tadpole
cancellation conditions  read 
\eqn\tadpole{\eqalign{
& \sum_{a=1}^K {{\rm N}_a\,  \prod_{I=1}^3{n_a^{I}}} = 16, \cr
& \sum_{a=1}^K {{\rm N}_a\,  n_a^{1} \prod_{I=2,3}
{\left(m_a^{I}+
       U_1^{I}\, n_a^{I}\right) }} = 0 , \cr
&  \sum_{a=1}^K {{\rm N}_a\,  n_a^{2} 
 \prod_{I=1,3}{\left(m_a^{I}+
        U_1^{I}\, n_a^{I}\right) }} =0 , \cr
&  \sum_{a=1}^K {{\rm N}_a\, n_a^{3} \prod_{I=1,2}{\left(m_a^{I}+
       U_1^{I}\, n_a^{I}\right) }} =0 .\cr}}
This can be compactly written as
\eqn\homo{
 \sum_{a=1}^K {\rm N}_a\ \Pi_a=\Pi_{\rm O6} }
where  $\Pi_a$ denotes the homological cycle of the wrapped D$6_a$-branes and
$\Pi_{\rm O6}$ the cycle of the orientifold planes along the $X_I$ axes
of  all three $T^2$s.  
In terms of the T-dual type I theory \tadpole\ refers to the cancellation of
the D9-brane and O9-plane charges, respectively the vanishing of the 
three possible types of D5-brane charges. 
The tree-level massless spectrum consists of ${\cal N}=4$ vectormultiplets
in the gauge group 
\eqn\gauge{ G=U(N_1)\times U(N_2)\times \ldots \times U(N_K) }
equipped with non-supersymmetric chiral matter in bifundamental, 
symmetric and antisymmetric representations of the gauge group\footnote{$^1$}{
More detailed information can be found in the papers
\refs{\rbgklnon,\rbgklmag,\rafiru,\rimr}.}.  
This chiral matter is localized at the intersections of
two D-branes and therefore each state appears with a multiplicity given by the
intersection number of the two D-branes. 
Taking these multiplicities into account, the 
RR-tadpole cancellation conditions guarantee the absence
of gauge anomalies in the effective four-dimensional low energy theories. 

The quite general form of the consistency conditions in terms
of the wrapping numbers of the D-branes allows for a bottom-up
approach to search systematically 
for  features of the Standard Model in the class of these 
intersecting brane models. In particular, in \rbkl\
a left-right symmetric model with three generations and gauge group
$SU(3)\times SU(2)_L\times SU(2)_R\times U(1)_{B-L}$
was constructed. Moreover, in \rimr\    
the matter content of a three generation $SU(3)\times SU(2)_L\times U(1)_Y$
Standard Model was found, containing also right handed neutrinos and
a slightly enlarged Higgs sector. In the first place the model 
had gauge group $U(3)\times U(2)\times U(1)\times U(1)$, but 
after analyzing mixed anomalies and the appropriate Green-Schwarz mechanism 
only the Standard Model gauge fields remained massless. The broken
gauge symmetries including lepton and baryon number survived as global
symmetries, thus guaranteeing the stability of the proton. 
Moreover, it was argued that at string tree level the radii of the three 
tori $T^2$ can be  tuned in such a way that
open string tachyons are absent. However, some of the tachyons
are welcome, as they can serve as Higgs bosons for breaking
the electroweak symmetry. 

Even though from a phenomenological 
point of view the models look quite interesting, we will show in the 
next section that the string theory is highly unstable.  

\newsec{NSNS tadpoles}
\smno
So far, for toroidal intersecting brane worlds  only the RR tadpole 
cancellation conditions were analyzed in detail. In this section
we will compute the NSNS tadpoles and derive the effective
scalar potential for the closed string moduli at open 
string tree level $e^{-\phi}$, which is next to leading order
in string perturbation theory. For the purpose of a phenomenological
application we perform the 
computation for four-dimensional models. 
 
The massless fields in the NSNS sector are the four-dimensional
dilaton and the 21 $\Omega{\cal R}$ invariant components of the
internal metric and the internal NS-NS two form flux.
In our factorized ansatz \split\ only 9 moduli are evident, which 
are the six radions $R_1^I$ and $R_2^I$ related to the size of the 
internal dimensions and the two-form flux $b^I_{12}$ on each $T^2$.
We extract the NSNS tadpoles from the
infrared divergences in the tree channel Klein-bottle, annulus and
M\"obius-strip amplitudes, the open string one-loop diagrams. 
Adding up the latter three contributions leads to a sum of 
perfect squares, from which we can read off the disc tadpoles.  
The computation is straightforward and the relevant formulas can be found in
\refs{\rbgklnon,\rbkl}. 
By adding up all three contributions we get for the dilaton tadpole
\eqn\dilaton{ \la \phi \ra_D=
 {1\over \sqrt{{\rm Vol}(T^6)}}\left( \sum_{a=1}^K  N_a\, 
    {\rm Vol}({\rm D}6_a) 
           -16\, {\rm Vol}({\rm O}6) \right) }
with 
\eqn\volumea{  {\rm Vol}({\rm D}6_a)=\prod_{I=1}^3 L^I({\rm D}6_a)=
        \prod_{I=1}^3 \sqrt{\left( n_a^I\, R_1^I\right)^2 + 
                          \left( (m_a^I+U_1^I\, n_a^I)\, R_2^I\right)^2} } 
and
\eqn\volumeb{  {\rm Vol}({\rm O}6)=\prod_{I=1}^3 L^I({\rm O}6)=
        \prod_{I=1}^3 R_1^I .} 
The result is simply the overall volume of the D6-branes and orientifold
planes, the latter ones  entering  with a negative sign. 
It is just the effective
four-dimensional tension in appropriate units.
Intriguingly, the dilaton tadpole can be expressed entirely in terms
of the complex structure moduli $U_2^I$ of the three $T_I^2$, 
\eqn\dilatonb{ \la \phi \ra_D= \left( \sum_{a=1}^K  N_a\, \prod_{I=1}^3 
   \sqrt{\left( n_a^I\, U_2^I\right)^2 + 
                          \left( (m_a^I+U_1^I\, n_a^I)\, 
                {1\over U_2^I}\right)^2}  
          -16\, \prod_{I=1}^3 U_2^I \right) .}
One way to understand this is to realize that the boundary and 
cross-cap states only couple to the left-right symmetric states of  
the closed string Hilbert space. The complex structure moduli 
are indeed left-right symmetric, whereas the K\"ahler
moduli appear in the left-right asymmetric sector, i.e. 
D-branes and orientifold O6-planes only couple to the complex
structure moduli. This is reversed in the T-dual type I picture, where the
tadpole only depends on the K\"ahler moduli. 

Besides the dilaton tadpole we also have three  
tadpoles for the imaginary parts of the complex structures, given by
\eqn\radion{ \la U^I_2 \ra_D=
 {1\over \sqrt{{\rm Vol}(T^6)}}\left( \sum_{a=1}^K  N_a\, 
           \Gamma^I({\rm D}6_a)\, L^J({\rm D}6_a)\, L^K({\rm D}6_a) 
           -16\, {\rm Vol}({\rm O}6) \right) }
with $I\ne J\ne K\ne I$ and 
\eqn\gam{  \Gamma^I({\rm D}6_a)={\left( n_a^I\, R_1^I\right)^2 - 
  \left( (m_a^I+ U_1^I\, n_a^I)\, R_2^I\right)^2\over  L^I({\rm D}6_a)} .} 
Analogous to \dilatonb\ these tadpoles can also be expressed
entirely in terms of the complex structure moduli $U_2^I$. 
Concerning type II models which have also been considered in similar
constructions \rafiru\ one needs to regard extra tadpoles for 
the real parts $U_1^I$, which cancel in type I. 
All NSNS tadpoles arise from the
following scalar potential  in string frame
\eqn\poten{ V(\phi,U_2^I)=  e^{-\phi}\, 
\left( \sum_{a=1}^K  N_a\, \prod_{I=1}^3 
   \sqrt{\left( n_a^I\, U_2^I\right)^2 + 
                          \left( (m_a^I+U_1^I\, n_a^I)\, 
                {1\over U_2^I}\right)^2}  
          -16\, \prod_{I=1}^3 U_2^I \right) }
with 
\eqn\potder{\la \phi \ra_D \sim {\partial V \over \partial \phi}, \quad\quad 
\la U^I_2 \ra_D \sim {\partial V \over \partial U^I_2} .}
The type II potential would only  change in erasing the term arising  from
the orientifold planes, and a third tadpole would appear due to $\la U_1^I
\ra_D\sim\partial V/\partial U_1^I$. 
Note that this potential is leading order in string
perturbation theory but already contains all higher powers in the complex
structure moduli, though we have only computed their one-point function
explicitly. One needs to be careful in interpreting it. 

In field
theory, the presence of a non-vanishing tadpole indicates that the tree-level
value was not chosen at a minimum of the potential. Even if one can compute
higher loop corrections formally, their meaning is very questionable, as we
expect fluctuations to be large no matter how small the coupling constant may
be. The theory is driven away to some distant minimum anyway, and 
perturbation theory around the unstable vacuum is impossible. 
As a second problem, the open string tachyons which are very often present in
non-supersymmetric string vacua would start to propagate at the open string 
loop level.\footnote{$^1$}{For a discussion of the stability regions of
intersecting D-branes with respect to the appearance of tachyons see \rrab.} 
Thus it is
mandatory to first shift to a minimum with vanishing tadpoles and without
tachyons before taking perturbations into account. 

In string theory the situation is even worse, as higher
loop corrections cannot even be computed because of infinities. 
If there appears a massless tadpole at genus $g$ in the string  
loop expansion,
one encounters a divergence at genus $2g$ from the region
in moduli space where a massless mode propagates
along a  long tube connecting two genus $g$ surfaces. 
So either one
finds a new vacuum by regarding the back-reaction of the massless fields along
the lines of \refs{\rfs,\rfstwo,\rdma,\rbfa}, 
or one uses a modification of the model where the tadpoles
are absent. We shall pursue the latter strategy in the following chapter.   

Actually, one could have anticipated the result \poten\ 
immediately, as the source
for the dilaton is just the tension of the branes, to first order given by
their volumes. The above expression is easily seen to arise from the
Dirac-Born-Infeld action for a D9$_a$-brane with constant $U(1)$ and
two-form flux
\eqn\borninfeld{
{\cal S}_{\rm DBI} = - 
T_p \int_{D9_a}{d^{10}x\ e^{-\phi} \sqrt{ {\rm det} \left
( G + ( F_a+B ) \right) } } }
including the D$p$-brane tension 
\eqn\tension{
T_p = {\sqrt{\pi} \over 16\kappa_0} \left( 4\pi^2 \alpha' \right)^{(11-p)/2}
.} 
One can take all background fields to be block-diagonal in terms of the
two-dimensional tori. There they take the constant values \rbgkl 
\eqn\constval{ 
G^{ij} = \delta^{ij}, \quad 
\left( F_a^I \right)^{ij} = {m_a^I \over n_a^I R_1^I R_2^I}
\epsilon^{ij},\quad 
\left( B^I \right)^{ij} = { b^I \over R_1^I R_2^I} \epsilon^{ij}\ {\rm with}\
b^I=0\ {\rm or}\ {1 \over 2} .}
Integrating out the internal six dimensions, regarding that the brane wraps
each torus $n_a^I$ times, one only needs to apply the T-duality to arrive at
\poten\ except for the negative contribution of the orientifold tension.
  
Due to the RR-tadpole cancellation condition and the triangle inequality,
the only point where all four tadpoles vanish is at
$U^I_2=\infty$. Interestingly, this proves the impossibility of a partial
breaking of supersymmetry in ${\cal N}=4$ vacua by relative angles
between D6-branes, respectively magnetic fluxes on
D9-branes.\footnote{$^2$}{This possibility has been established in 
${\cal N}=2$ type II and heterotic vacua under certain rather special 
conditions
\refs{\rapt,\rfgptwo,\rtv,\rcklt}.}   

The potential displays the usual runaway behaviour
one often encounters in non-supersymmetric string models. 
The complex structure is dynamically pushed to the degenerate limit,
where all branes lie along the $X_I$ axes and the $Y_I$ directions
shrink, keeping the volume fixed. Put differently, the positive tension of the
branes pulls the tori towards the $X_I$-axes. 
The typical runaway slope being set by the tension \tension\ proportional to
the string scale, a `slow rolling' does not appear to be feasible either. 
Apparently, this has dramatic consequences for all toroidal intersecting
brane world models. 
They usually require a tuning of parameters at tree-level and  
assume the global stability of the background geometry as given by the closed
string moduli. 
If at closed string tree-level one has arranged the radii
of the torus such, that open strings stretched between D-branes
at angles are free of tachyons, dynamically the system flows towards
larger complex structure and will eventually reach a point where
certain scalar fields become tachyonic and indicate a decay of
the brane configuration. 

Via T-duality the instability translates back into a dynamical
decompactification towards the ten-dimensional supersymmetric vacuum. 
Thus, even if from a heuristic point of view toroidal intersecting 
brane world models look quite promising, the non-supersymmetric
string theory is highly unstable. 

\newsec{Orbifold intersecting brane models}
\noindent
One  way to avoid this runaway behaviour of the complex structure 
moduli is to freeze them from the very beginning. This can be 
achieved by  
dividing the toroidal model by an appropriate discrete symmetry. 
For instance, for the left-right symmetric orbifold $\ZZ_3$ 
acting as
\eqn\zdrei{   \Theta:  Z^I\to e^{2\pi i/ 3}\, Z^I }
on all three complex coordinates, the complex structure
on all three $T^2$'s is fixed to be either\footnote{$^1$}{See \rba\ for a
discussion of discrete parameters in type I vacua.} 
\eqn\comfix{   U^I_{\bf A}={1\over 2} +i {\sqrt{3}\over 2} }
or
\eqn\comfixb{   U^I_{\bf B}={1\over 2} +i {1\over 2\sqrt{3}} .}
In \refs{\rbgka,\rbgkc}, 
where this type of orbifold was considered for the first time,  
the torus \comfix\ with K\"ahler modulus 
\eqn\kaehlera{T^I_{\bf A} = i{\sqrt{3} \over 2} R^2 }
was called the {\bf A}-torus and the
torus \comfixb\ with 
\eqn\kaehlerb{T^I_{\bf B} = i{1 \over 2\sqrt{3}} R^2 }
the {\bf B}-torus. Note, that under T-duality
these models are mapped to asymmetric type I orbifolds where the
K\"ahler moduli are frozen. 
Vice versa, left-right symmetric type I orbifolds are mapped
to asymmetric $\Omega{\cal R}$ orientifolds.  
Therefore, only for left-right symmetric $\Omega{\cal R}$ 
orientifolds with intersecting branes 
the disc scalar potential does not depend on the $U^I$, 
preventing the torus from shrinking to degenerate limits. 

Thus, we are naturally led to consider the orientifold
\eqn\typeone{  {{\rm Type\ IIA}\ {\rm on}\ T^{6} \over 
                  \{\ZZ_3+\Omega{\cal R} \ZZ_3\} } .}
This is precisely one example of the supersymmetric 
orientifolds with D6-branes at angles
introduced in \refs{\rbgka,\rbgkc,\rprada,\rbgkb,\rpradb}\footnote{$^1$}{In \rfhs\ 
extensions to 
$\ZZ_N \times \ZZ_M$ orbifold groups have been considered.}. 
In the closed string sector this Z-orbifold has Hodge numbers
$(n_{21},n_{11})=(0,36)$, where 9 K\"ahler deformations
come from the untwisted sector and the remaining 27 are the
blown up modes of the fixed points. As noted before, this manifold
has frozen complex structure. 
Due to the $\ZZ_3$ symmetry we have three kinds 
of O6-planes located as indicated in figure 2, being identified under the
orbifold action.  
\fig{}{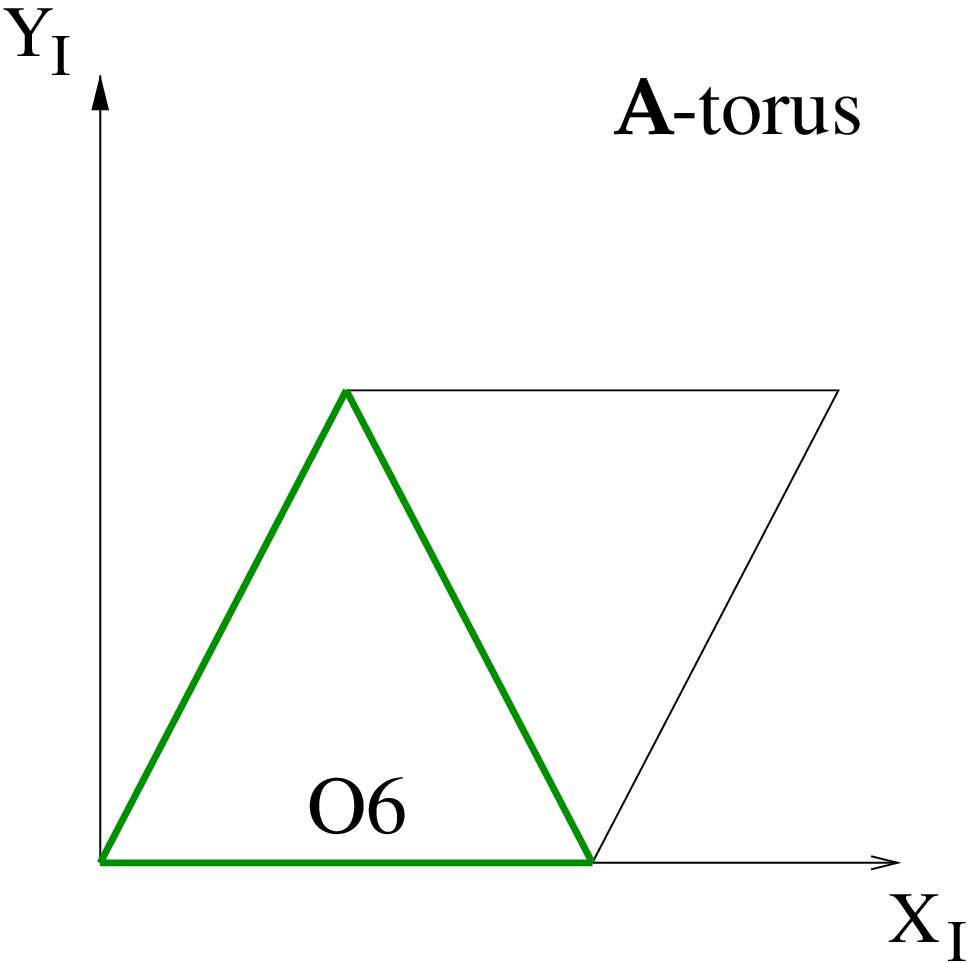}{6truecm}
One can cancel the Klein bottle tadpoles locally by introducing
four D6-branes on top of the O6-planes leading to a supersymmetric model
with gauge group of rank $2=16\cdot 2^{-3}$, 
in accord with the rank reduction normally
encountered in type I vacua with NSNS $B$-field of rank 6. 
In the following we discuss the more general case, where one introduces
D6-branes intersecting at angles into such a background and in particular
determine the tadpole cancellation conditions and the chiral
massless spectrum. In particular, we find that the phenomenological obstruction
of the small rank can be lifted when putting branes at arbitrary angles on the
orbifold space. In the dual flux picture this means that adding magnetic flux
to the supersymmetric theory allows to have a larger gauge symmetry. This is
very surprising in the first place, one would have naively expected the
opposite to happen. 
But, effectively, a D9-brane with additional flux can carry
less RR charge than without. 

Note the difference compared to \rfhstwo\ where 
six-dimensional supersymmetric orientifolds of
type I$'$ have been combined with generic brane
configurations on an extra $T^2$. This latter compactification indeed suffers
from the very small rank of the gauge group. 

An individual D6$_a$-brane is again 
determined by three pairs of wrapping numbers
$(n^I,m^I)$ along the fundamental cycles 
\eqn\cycles{
e_1^{\bf A} = e_1^{\bf B} = R , \quad 
e_2^{\bf A} = {R \over 2} + i {\sqrt{3} R \over 2}, \quad 
e_2^{\bf B} = {R \over 2} + i {R \over 2\sqrt{3} } }
of each $T^2$. 
Under the $\ZZ_3$ and $\Omega{\cal R}$ symmetry in general
the  branes are organized in orbits of  length six. 
Such an orbit constitutes an equivalence  class $[a]$ 
of D6$_a$-branes denoted by $[(n_a^I,m_a^I)]$. 
For the {\bf A}-torus the six branes contained in the equivalence  class
$[(n^I,m^I)]$ are given by
\eqn\equiva{ \matrix{ & {\ds {n^I \choose m^I}} & 
          {\buildrel \ZZ_3 \over \Rightarrow} &
             {\ds {-n^I-m^I \choose n^I}} & {\buildrel \ZZ_3 \over \Rightarrow} &
             {\ds {m^I \choose -n^I-m^I}} \cr
              \Omega{\cal R} &     \Downarrow & &
                 \Downarrow & &
                 \Downarrow \cr
           & {\ds {n^I+m^I \choose -m^I}} & {\buildrel \ZZ_3 \over \Leftarrow} &
             {\ds {-m^I \choose -n^I}} & {\buildrel \ZZ_3 \over \Leftarrow} &
             {\ds {-n^I \choose n^I+m^I}} \cr  } }
and for the {\bf B}-torus by
\eqn\equiva{ \matrix{ & {\ds {n^I \choose m^I}} & 
           {\buildrel \ZZ_3 \over \Rightarrow} &
             {\ds {-2n^I-m^I \choose 3n^I+m^I}} & 
             {\buildrel \ZZ_3 \over \Rightarrow} &
                {\ds {n^I+m^I \choose -3n^I-2m^I}} \cr
              \Omega{\cal R} &     \Downarrow & &
                 \Downarrow & &
                 \Downarrow \cr
         & {\ds {n^I+m^I \choose -m^I}} & {\buildrel \ZZ_3 \over \Leftarrow} &
           {\ds {n^I \choose -3n^I-m^I}} & {\buildrel \ZZ_3 \over \Leftarrow} &
             {\ds {-2n^I-m^I \choose 3n^I+2m^I}}. \cr  } }
As an example for an orbit on 
a single {\bf A}-torus the equivalence class $[(2,1)]$ is shown in figure 3. 
The solid lines represent the images under $\ZZ_3$ and the dashed lines
the $\Omega{\cal R}$ mirror branes. Due to the relation
\fig{}{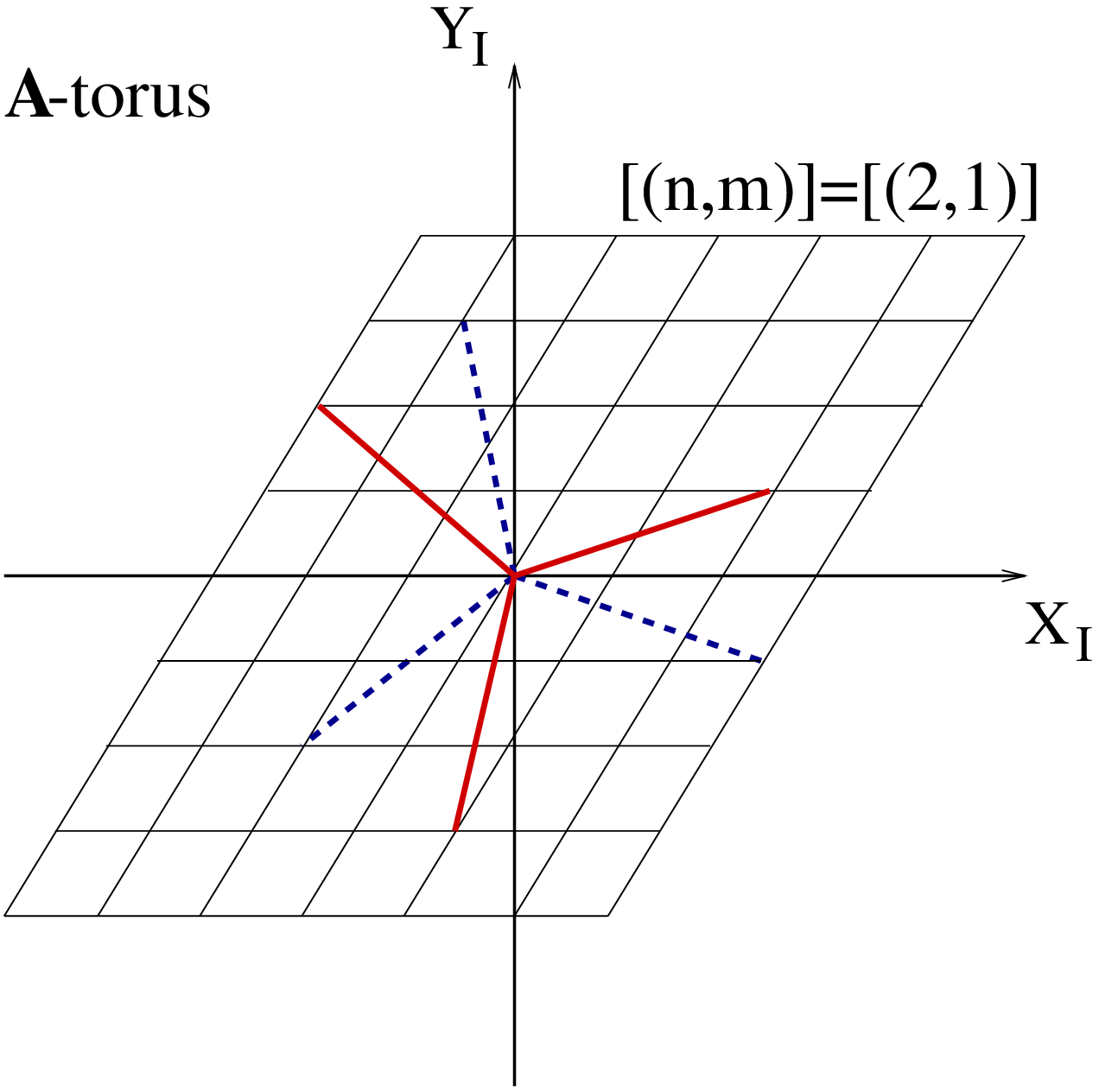}{8truecm}
\eqn\noncom{   \Theta\, (\Omega{\cal R}) = (\Omega{\cal R})\,  \Theta^{-1} }
only untwisted sector fields couple to the orientifold planes. 
This is also clear from the fact that the orientifold planes are of
codimension one on each $T^2$ and therefore can avoid a blown-up
$\IP^1$ from an orbifold fixed points. Similarly, the
D6-branes can not wrap around the blown-up cycles to become 
fractional branes and thus are not charged under the twisted sector
RR-fields. Thus, there are only untwisted tadpoles.  

\subsec{Tadpoles}
\noindent
Combining the results from \refs{\rbgklnon,\rbgkc} 
the computation of the Klein-bottle, annulus
and M\"obius strip amplitudes is a straightforward exercise and we will
only present the salient features and results of this rather tedious
computation. 
Since the complex structure is fixed we only get one RR and one NSNS
tadpole cancellation condition. In the annulus amplitude  all open string
sectors contribute including those from open strings stretched between 
two branes belonging to the same equivalence class. It turns out to 
be convenient to define the following two quantities for any equivalence class
$[(n^I_a,m^I_a)]$ of D6$_a$-branes 
\eqn\yizi{\eqalign{
       &Z_{[a]}={2 \over 3} \sum_{({n}^I_b,{m}^I_b)\in [a]}
                           \prod_{I=1}^3  \left( {n}^I_b +{1\over 2}\,
                         {m}^I_b \right) , \cr
       &Y_{[a]}= -{1\over 2} \sum_{({n}^I_b,{m}^I_b)\in [a]}
                          (-1)^M\, \prod_{I=1}^3  
                         {m}^I_b  \cr }, }
where $M$ is defined to be odd for a mirror brane and otherwise even. The sums 
are taken over all the individual D6$_b$-branes that are elements of the
orbit $[a]$.       
The explicit expressions for $Z_{[a]}$ and $Y_{[a]}$ for the
four possible tori, {\bf AAA, AAB, ABB, BBB}, 
can be found in appendix A. 
If we introduce $K$ stacks of equivalences classes $[a]$  of branes,
then the RR-tadpole cancellation condition reads
\eqn\tadpolec{    \sum_{a=1}^K N_a\, Z_{[a]} =2 .}
Note, that the sum is over equivalence classes of D6-branes.
In fact, $Z_{[a]}$ is the projection of the entire orbit of D6$_a$-branes onto
the $X_I$ axes, i.e. the sum of their RR charges with respect to the dual
D9-brane charge. Therefore the
appearance of $Z_{[a]}$ in the tadpole cancellation condition is very natural,
simply meaning that the RR-charges of all D6-branes have to cancel
the RR-charges of the orientifold O6-planes. 
If the $Z_{[a]}$ are all positive, as was the case in the supersymmetric 
solutions of \refs{\rbgka,\rbgkc}, then \tadpolec\ implies a very small rank of
the  gauge group. However, in the general case the $Z_{[a]}$ may also be
negative so that also gauge groups of higher rank can be realized. 

In the closed string NSNS sector all scalars related to the complex
structure moduli are projected out under $\ZZ_3$, so that only the dilaton
itself can have a disc tadpole. This is indeed what we find from the
tree-channel one loop amplitudes, as the only divergences there
comes from the dilaton. The scalar potential for our model is
\eqn\potzdr{  V(\phi)=e^{-\phi} \left( \sum_{a} N_a \prod_{I=1}^3  L_{[a]}^I -2
                         \right) }
with the lengths given by
\eqn\lls{
          L_{[a]}^I=\cases{ \sqrt{ (n_a^I)^2 + (m_a^I)^2 + n_a^I m_a^I } &
         {\rm for\ the\ {\bf A}-torus} , \cr 
            & \phantom{dfdf} \cr
           \sqrt{ (n_a^I)^2 + {1\over 3} (m_a^I)^2 + n_a^I m_a^I } &
         {\rm for\ the\ {\bf B}-torus} .\cr}} 
Thus, similar to the toroidal case discussed in section 2, whenever 
the D-branes do not lie on top of the orientifold planes the dilaton
tadpole does not vanish. 
In this way the {\it local} cancellation of the RR charge is in one to one
correspondence with supersymmetric vacua and the cancellation of NSNS
tadpoles. The only exception to this rule appears to be a parallel
displacement of orientifold planes and D-branes, i.e. a Higgs mechanism
breaking $SO(2N_a)$ to $U(N_a)$.    

\subsec{Massless spectrum}
\noindent
Having found the one-loop consistency condition the next step is to determine
the massless spectrum and to see whether one can find 
phenomenologically interesting models. 
In the closed string sector at string tree level ${\cal N}=1$ supersymmetry
is preserved and we get the same massless spectrum of vector and
chiral multiplets as in \rbgkc.
However, in the open string sector we break supersymmetry
and get  more interesting spectra.
For the supersymmetric brane configurations the massless spectra for
these kinds of orientifolds with D-branes at angles were always 
non-chiral, which is no longer true in the non-supersymmetric case. 

In the following we discuss the most generic situation where all equivalence 
classes contain six different D6-branes, i.e. there are no dual D9-branes
without any magnetic flux on their world volume.
A string with both ends on the same individual brane in some 
equivalences class $[a]$ 
gives rise to an ${\cal N}=4$ vectormultiplet in the gauge group $U(N_a)$. 
Open strings stretched  between branes belonging to two different classes can 
break supersymmetry and give rise to chiral fermions in the
bifundamental representations of the gauge groups. There
are $36=6\times 6$ different open string sectors of this kind. 
Due to the $\ZZ_3$ and $\Omega{\cal R}$ symmetry
only $6$ of them are independent. Thus, we can pick one brane, D$6_a$,
from the first stack  and determine the massless spectrum with all
$6$ branes D$6_{bi}$, $i\in\{1,\ldots,6\}$,  from the second stack. 
Open strings between    
D$6_a$ and $\ZZ_3$ images of D$6_{b}$, i.e. $i=\in\{1,2,3\}$, 
yield chiral fermions
in the $(\o{N}_a,N_b)$ representation and open strings between    
D$6_a$ and mirror images,  i.e. $i=\in\{4,5,6\}$,  
in the second stack give rise to chiral fermions
in the $({N}_a,N_b)$ representation. The multiplicity of these
massless states is determined by the topological intersection number
between the branes in question, where intersections with formally negative
intersection number have flipped orientation 
leading to the conjugate representations.

In the end, only the net number of such fermion generations is relevant. For
instance, let D6$_{b1}$ and D6$_{b2}$ be two different branes in the orbit
$[b]$ and D6$_a$ another one in the orbit $[a]$. 
Assume, that in the D$6_a$-D$6_{b1}$ sector we  have a chiral 
fermion $\psi_{a,b1}$ 
in the $(\o{N}_a,N_b)$ representation and in  the  
D$6_a$-D$6_{b2}$ sector a fermion $\psi_{a,b2}$ in the conjugate 
$({N}_a,\o{N}_b)$ representation. In principle, these 
can pair up to yield a Dirac mass term with mass of the order of 
the string scale.  
Indeed, since in the D$6_{b1}$-D$6_{b2}$ sector
we get a massless scalar $H_{b1,b2}$
in the adjoint representation of $U(N_b)$, the three-point
coupling on the disc diagram as shown in figure 4 exists.  
\fig{}{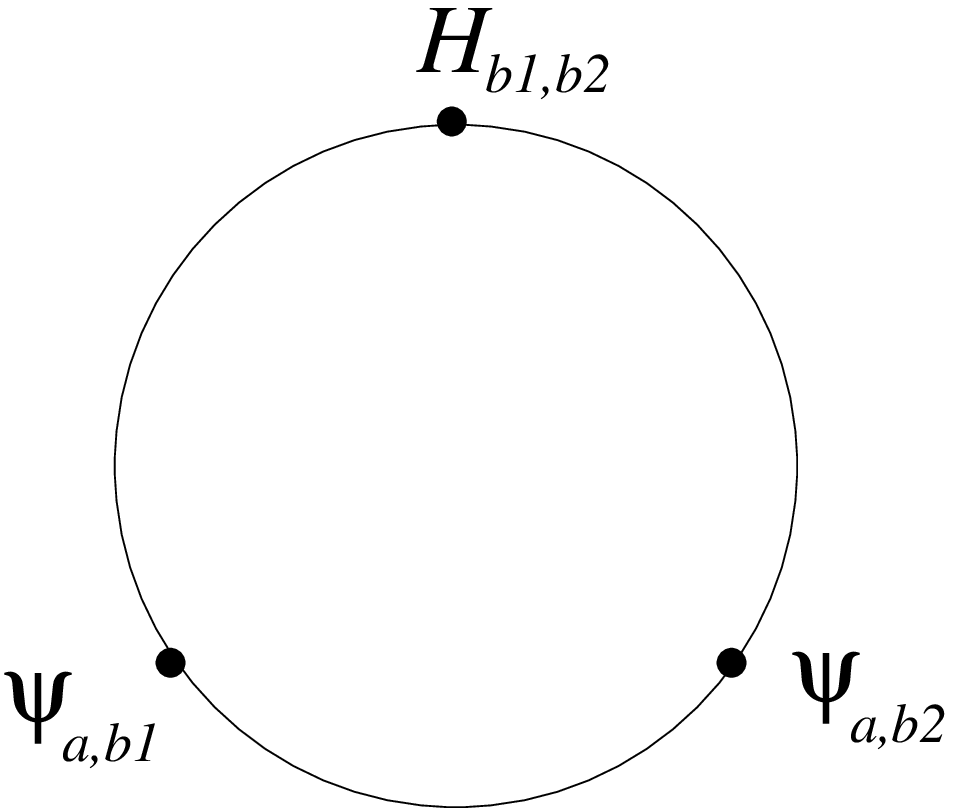}{6truecm}
\noindent
Giving a vacuum expectation value to the $SU(N_b)$ singlet in 
the adjoint of $U(N_b)$ leaves the gauge symmetry unbroken and gives a mass
to the fermions. From the string point of view, this deformation
is exactly the one studied in \rbbh, which deforms the two intersecting
branes of the orbit $[b]$ into a single brane wrapping
a supersymmetric cycle.  

For the relevant net number of chiral left-handed bifundamentals one 
obtains the  following simple expressions
\eqn\bifun{\eqalign{
         &(\o{N}_a,N_b)_L:\quad    Z_{[a]}\, Y_{[b]}- Y_{[a]}\, Z_{[b]} , \cr
          &({N}_a,N_b)_L:\quad     Z_{[a]}\, Y_{[b]}+ Y_{[a]}\, Z_{[b]} .\cr }}
Thus, the combinations $Z_{[a]}$ and $Y_{[a]}$ can be interpreted
as effective wrapping numbers. 

Finally, we have the open strings stretched between different branes of the
same equivalence class. Open string between the brane D$6_a$ and the
three images under $\Omega{\cal R}\Theta^k$ 
give rise to chiral fields in the symmetric
and antisymmetric representation. The net numbers of these
massless fields are given by
\eqn\asymm{\eqalign{
         (A_a)_L:&\quad    Y_{[a]}  , \cr
        (A_a+S_a)_L:&\quad     Y_{[a]}\left( Z_{[a]}-{1\over 2}\right) .\cr }}
Finally, open strings between the brane $D6_a$ and its two $\ZZ_3$ images
yield massless fermions in the adjoint representation
\eqn\adj{ ({\rm Adj})_L:\quad  3^{n_{\bf B}} \prod_{I=1}^3 
          \left(L^I_{[a]}\right)^2 ,}
where 
$n_{\bf B}$ counts the number of {\bf B}-tori in $T^6$. This latter sector
is ${\cal N}=1$ supersymmetric, as the $\ZZ_3$ rotation alone preserves
supersymmetry.   
Before going into the phenomenological details we first discuss the issue of 
gauge anomalies in the next section. 

\subsec{Anomaly cancellation}
\noindent
Since we have chiral fermions, there are potential gauge anomalies,
which  however should be absent due to the string theoretic one loop
consistency of our models. 
For the spectrum shown in section (4.2) we obtain that the non-abelian
gauge anomaly of the  $SU(N_a)$ gauge factor is proportional to
\eqn\gaugeano{ \sum_{b\ne a} 2\, N_b\, Z_{[b]}\, Y_{[a]} + (N_a-4)\, Y_{[a]} +
                2\, N_a\, Y_{[a]} \left(Z_{[a]}-{1\over 2}\right) ,}
which vanishes when we use the RR-tadpole cancellation condition \tadpolec\
in  the first term in \gaugeano.
As usual, the abelian gauge anomalies do
not cancel right away. 
Indeed the $U(1)_a-g_{\mu\nu}^2$ anomalies are proportional to
\eqn\abelan{      3\, N_a\, Y_{[a]} }
and the mixed $U(1)_a-U(1)_b^2$ anomalies are 
\eqn\abelanb{      2\, N_a\, N_b\, Y_{[a]} Z_{[b]}.}
In order to cancel these anomalies one has to invoke a generalized
Green-Schwarz mechanism. It was pointed out in \rafiru\ that the relevant
axions are among the untwisted sector RR-fields. 
Using the same notation, we are discussing the couplings in the
T-dual type I language where angles are translated into fluxes.
In ten space-time dimensions we have the RR fields $C_2$ and
$C_6$, $dC_6=*dC_2$, with world-volume couplings
\eqn\wwc{ \int_{D9_a} C_6\wedge  F_a\wedge F_a, \quad\quad  
             \int_{D9_a} C_2\wedge F_a\wedge F_a\wedge F_a\wedge F_a. }
Upon dimensional reduction to four dimensions we only get one
two-form, $B_2^0=C_2$. Note, that the other three type I two-forms
\eqn\postwo{ B^I_2=\int_{T^2_J\times T^2_K} C_6 }
are projected out by the $\ZZ_3$ symmetry. 
The dual four-dimensional axion $C_0^0$ is given by
\eqn\posdual{ C^0_0=\int_{T^2_1\times T^2_2\times T^2_3} C_6 .}
From the ten dimensional couplings \wwc\ summed over an entire orbit of the
symmetry group one obtains the
four-dimensional couplings of the RR-forms to
the gauge fields
\eqn\green{\eqalign{   &N_a\, Y_{[a]} \int_{M_4} B^0_2\wedge F_a, \cr 
                     &N_b\, Z_{[b]} \int_{M_4} C^0_0\, F_b\wedge F_b. \cr}}
Apparently, these couplings have precisely the right form
to cancel the abelian gauge anomalies \abelan\ and \abelanb\ via a 
generalized Green-Schwarz mechanism.
In fact there is only one anomalous $U(1)$ 
\eqn\massiv{    F_{\rm mass}=\sum_a  (N_a\, Y_{[a]})\, F_a}
which becomes massive due to the first coupling in \green. 
It can be checked that the mixed $U(1)-G^2$ anomalies are also
canceled by this generalized Green-Schwarz mechanism.

\subsec{Stability}
\noindent
As the tadpole calculation \potzdr\  
shows, after projecting out the complex structure
moduli, the potential is flat for the remaining moduli at 
the leading order in the string perturbation theory. 
Only the dilaton still runs away to zero coupling. 
For the background geometry, we certainly expect to find corrections in higher
loop diagrams, when untwisted and twisted K\"ahler moduli, 
decoupled only at leading order,  can run around the loop. 

Definite statements can hardly be made about the higher loop
potentials, but at least qualitatively we can say something
about the one-loop potential for the untwisted K\"ahler moduli
$K_U^I=R_1^I\, R_2^I$. 
Since the closed string sector is supersymmetric the only
dependence of the one-loop potential on $K_U^I$ can arise
from the annulus and the M\"obius strip amplitudes. 
Since we assumed that no D6-branes lie along the
$X_I$ axes, the massless strings in these sectors are localized at the
intersection points of the respective D6-branes and O6-planes. 
They do not `see' the global geometry of the torus and, hence, 
there is no explicit dependence of the M\"obius strip
amplitude on the radii. 

Therefore, the only contribution can come
from non-supersymmetric sectors in  the annulus amplitude,
which depend on the radii. These arise from open strings stretched
between two intersecting D6-branes, which are parallel on one or two 
tori. In the directions of the latter tori there are both Kaluza-Klein
and winding modes for the open strings. Since the KK modes scale
like $1/(R^I)^2\sim 1/K_U^I$ and the winding modes like $(R^I)^2\sim K_U^I$ 
there is a good chance
that the one-loop scalar potential stabilizes the untwisted
K\"ahler moduli. 
Of course, in this argument we have assumed that the Fischler-Susskind
mechanism does not qualitatively change the picture we derived
from the flat tree-level background. Alternatively one can also proceed
in a way analogous to the tadpoles associated to the complex structures
$U^I$, namely considering orbifold groups where all radii are frozen, like
the $\ZZ_3\times\widehat{\ZZ}_3$ orientifold of \refs{\rbimp,\rbgkl}.


Another issue concerns the existence of open string tachyons, which also may
spoil stability at the open string loop-level. 
In general, the bosons of lowest energy in a non-supersymmetric open 
string sector can have negative mass squared. Here one has to distinguish 
two different cases. Either the two D-branes in question
intersect under a non-trivial angle on all three two-dimensional  tori or 
the D-branes are parallel on at least one of the tori $T^2_I$. 
In the latter case one can get rid of the tachyons at least classically 
by making the distance between the two 
D-branes on the torus $T^2_I$ large enough. 
In the former case, it depends on the three angles, $\varphi_{ab}^I$,  between
the  branes D$6_a$ and D$6_b$  whether there appear tachyons or not.  

Defining $\epsilon_{ab}^I=\varphi_{ab}^I/\pi$
and let $P_{ab}$ be the number of  $\epsilon_{ab}^I$ satisfying
$\epsilon_{ab}^I>1/2$,
to compute the ground state energy in this twisted
open string sector one has to distinguish the following three cases
\eqn\energy{ E^0_{ab}=\cases{
                {1\over 2} \sum_I |\epsilon_{ab}^I|-{\rm max}\left\{
                      |\epsilon_{ab}^I| \right\} &
                  for $P_{ab}=0,1$ \cr
           1+{1\over 2}\left( |\epsilon_{ab}^I|-|\epsilon_{ab}^J|-
           |\epsilon_{ab}^K|\right) & \cr 
            \quad  -
           {\rm max}\left\{ |\epsilon_{ab}^I|,1-|\epsilon_{ab}^J|,
                  1-|\epsilon_{ab}^K| \right\} &
                  for $P_{ab}=2$ and  $|\epsilon_{ab}^I|\le{1\over 2}$\cr
           1-{1\over 2} \sum_I |\epsilon_{ab}^I| & 
              for $P_{ab}=3$. \cr }}
In order for a brane model  to be free of tachyons, for all 
open string sectors $E^0_{ab}\ge 0$ has to be satisfied. Since in the orbifold
model each brane comes with a whole equivalence class of branes,  
and the angles between two branes do not depend on any moduli (like
in the toroidal case), freedom of tachyons  is quite a strong condition.
We will discuss this point further for the concrete three generation
models in section 5. We shall actually find that, even though tree-level
stability is a strong condition, it can be satisfied in particular 
cases. 

Even if classically we can avoid tachyons by moving parallel branes 
far apart, at quantum level effective potentials for these 
open string moduli are generated which might spoil the stability 
of the configuration by pulling the branes together until
a tachyon reappears. Without knowing the precise scalar potential
definite statements can not be made. 

\newsec{Three generation models}
\noindent
In this section we will try to solve the consistency equations 
for the $\ZZ_3$ orbifold to search for   models which come as
close as possible to the Standard Model, respectively a moderate extension. 
Amazingly, even in this fairly constrained orbifold set-up
it is not too difficult to get three 
generation models with $SU(3)\times SU(2)_L\times U(1)_Y\times U(1)_{B-L}$
or $SU(5)\times U(1)$ gauge group and Standard Model matter fields enhanced
by  a right-handed neutrino. Only when it comes to the Higgs sector
and the Yukawa couplings we encounter some deviations
from our Standard Model expectations.

In \refs{\rafiru,\rbkl,\rimr} three generation intersecting brane worlds 
were always realized on four stacks of D6-branes
with gauge group  $U(3)\times U(2)\times U(1)\times U(1)$ and chiral
matter only in the bifundamental representations of the gauge factors.
It turns out that such a scenario is not possible in our
orbifold case for the following reason. Requiring that there does not
exist any matter in the antisymmetric representation of $U(3)$ forces
$Y_1=0$. Due to \bifun, this in turn implies that we get the same 
number of chiral fermions in the $({\bf 3,2})$ and in the $({\bf \o 3, 2})$ 
representation of $U(3)\times U(2)$ leading
to an even number of left-handed quarks. Thus, employing only 
bifundamental fields is not sufficient. 

\subsec{Extended Standard Model}
\noindent
We are forced to realize the right-handed $(u,c,t)$-quarks 
in the antisymmetric representation of $U(3)$, which, accidentally, is
the same as the anti-fundamental representation $\bf\o 3$. 
Moreover, requiring that there does not appear any chiral matter in the
symmetric representation of $U(3)$ and $U(2)$ forces us to have
$Z_3=Z_2=1/2$. After some inspection one realizes that the best way to approach
the Standard Model is to start with only three stacks of D6-branes 
with gauge group $U(3)\times U(2)\times U(1)$ and
\eqn\threegen{ (Y_1,Z_1)=\left(3,{1\over 2}\right), \quad
               (Y_2,Z_2)=\left(3,{1\over 2}\right), \quad
               (Y_3,Z_3)=\left(3,-{1\over 2}\right) .}
Note, that this choice indeed satisfies the RR-tadpole cancellation
condition \tadpolec. 
The three generation chiral massless spectrum is shown in table 4. 
\vskip 0.8cm
\vbox{
\centerline{\vbox{
\hbox{\vbox{\offinterlineskip
\def\tablespace{height2pt&\omit&&\omit&&\omit&&
 \omit&\cr}
\def\tablerule{\tablespace\noalign{\hrule}\tablespace}

\hrule\halign{&\vrule#&\strut\hskip0.2cm\hfill #\hfill\hskip0.2cm\cr
\tablespace
& matter  && $SU(3)\times SU(2)\times U(1)^3$ && $U(1)_Y$ && $U(1)_{B-L}$ &\cr
\tablerule
& $\left( Q_L\right)_i$ && $({\bf 3},{\bf 2})_{(1,1,0)}$ &&  ${1\over 3}$ && 
${1\over 3}$ & \cr
\tablespace
&  $\left(u^c_L\right)_i$ && $(\o{\bf 3},{\bf 1})_{(2,0,0)}$ && 
        $-{4\over 3}$ &&  $-{1\over 3}$  & \cr
\tablespace
& $\left(d^c_L\right)_i$ && $ (\o{\bf 3},{\bf 1})_{(-1,0, 1)}$ && 
       ${2\over 3}$ && $-{1\over 3}$   & \cr
\tablerule
& $\left(l_L\right)_i$ && $({\bf 1},{\bf 2})_{(0,-1,1)}$ && ${-1}$ && 
           ${-1}$   & \cr
\tablespace
& $\left(e^+_L\right)_i$ && $({\bf 1},{\bf 1})_{(0,2,0)}$ && ${2}$ && 
            ${1}$   & \cr
\tablespace
& $\left(\nu^c_L\right)_i$ && $({\bf 1},{\bf 1})_{(0,0,-2)}$ && ${0}$ && 
                  ${1}$    & \cr
\tablespace}\hrule}}}}
\centerline{
\hbox{{\bf Table 4:}{\it ~~ Left-handed fermions for the 3 generation 
model.}}}
}
\vskip 0.5cm 
\noindent
Of course, we now assume that the non-chiral fermions have 
paired up and decoupled  as was described earlier.
Note, that the right-handed leptons are realized as open strings in the
antisymmetric representation of $U(2)$ and the right-handed neutrinos
as open strings in the symmetric representation $\o{\bf S}$
of the $U(1)$ living
one the third stack of D6-branes.   
As expected from the general analyses of the $U(1)$ anomalies 
there is one anomalous $U(1)$  gauge symmetry
\eqn\anomalous{    U(1)_{\rm mass}=3\, U(1)_1 +2\, U(1)_2 + U(1)_3 }
and two anomaly free ones which can be chosen to be $U(1)_Y$ and
$U(1)_{B-L}$
\eqn\anomalous{ \eqalign{   U(1)_Y&=-{2\over 3}\, U(1)_1 + U(1)_2 , \cr
                            U(1)_{B-L}&=-{1\over 6}\left( U(1)_1 -3\, U(1)_2
                 +3\, U(1)_3\right). \cr  }}
Analogous to \refs{\rafiru,\rbkl,\rimr}, 
since the one-loop consistency of the string model 
requires  the formal cancellation 
of the $U(2)$ and $U(1)$ (non-abelian) gauge anomalies, 
the possible models are fairly constrained and require the introduction
of right-handed neutrinos. 
Because  the lepton number is not a global symmetry of the model, 
there exists the possibility to obtain Majorana mass terms and invoke  
the see-saw mechanism for the neutrinos mass hierarchy. 
Since, after introducing the right-handed
neutrino into the Standard Model, the $U(1)_{B-L}$ symmetry becomes
anomaly-free, it is not too surprising that in the string theory
this symmetry is gauged. After the Green-Schwarz mechanism the anomalous
$U(1)_{\rm mass}$ decouples, but survives as a global symmetry. 
Thus, the possible Yukawa couplings are more constrained than in the
Standard Model. 

It is straightforward but extremely tedious to find realizations of the
$(Y_{a},Z_{a})$ given above in terms of actual winding numbers
$[(n_{a}^I,m_{a}^I)]$. We have performed a systematic computer search and
identified 36 solutions for each stack $[a]$, the number being
independent of which of the four possible types of the torus had been chosen. 
Surprisingly, all winding numbers range
between -3 and 3, and only for the {\bf BBB} torus from $-5$ and $5$. 
The actual number of inequivalent string models with the
above mentioned Standard Model like features then is $4\cdot 36^3$.  

\subsec{Stability and Higgs scalars}
\noindent
As mentioned already, the primary motivation to study the present class of
$\Omega{\cal R}$ orientifolds of type IIA is their stability. 
While the closed string sector does not suffer from any massless disc
tadpole apart from the
dilaton, and thus all moduli sit at extrema of their potential, the open
string sector contains tachyonic scalars which indicate an instability.
Assuming the closed string moduli to be qualitatively unaffected by the
condensation, 
the endpoint of this will presumably be a new vacuum, where the isolated
cycles themselves are general supersymmetric ones 
but still intersects each other in a non-supersymmetric way. 
Still everything will be stable with respect to tachyons. 
There are two different 
patterns of gauge symmetry breaking which arise when any two branes
condense via this mechanism. 
When the two branes are of different class $[a]$ and $[b]$, 
the tachyonic Higgs field is in the bifundamental representation of the
$U(N_{a}) \times U(N_{b})$ gauge group and the condensation resembles the
Higgs mechanism of electroweak symmetry breaking. On the contrary, when the
two branes are elements of the same orbit $[a]$, the Higgs field will be in
the antisymmetric, symmetric or adjoint representation of the  $U(N_{a})$
and thus affect only this factor.  

The version of the Standard Model extended by a gauged $B-L$ symmetry together
with right-handed neutrinos requires a two step gauge symmetry breaking (For
the probably first appearance of such models see \rcgpr.). In
order to avoid conflicts with various experimental facts a hierarchy of Higgs
vacuum expectation values is required. First the $U(1)_{B-L}$ has to be broken
at a scale at least some $10^{4-6}$ above the electroweak scale. This requires
a Higgs field charged under this group but a singlet otherwise, which can be
met with a tachyon from a sector of strings stretching between two branes in
the orbit that supports the $U(1)_3$. 
The second step is the familiar electroweak symmetry breaking
which needs a bifundamental Higgs doublet. 

We have therefore performed a study among all the $4\cdot 36^3$ models looking
for such a  suitable tachyon spectrum. 
In any sector of open strings stretching between two D6-branes $a$ and
$b$ the lightest physical state has a mass given by \energy. 
By expressing the angle variables in terms of winding numbers, one can 
set up a computer program to do the search for models with a Higgs scalar in
the $({\bf 2,1})$ and/or another one in the `symmetric' representation of
$U(1)_{B-L}$. All other open string sectors need to be free of tachyons. 
The results are the following: 
For the {\bf AAA} and the {\bf BBB} type tori one can get  D6-brane
configurations that display only tachyons charged under $U(1)_{B-L}$, but none
of these models does have a suitable Higgs in the $({\bf 2,1})$. Vice versa,
the {\bf AAB} and {\bf ABB} models do have Higgs fields in the $({\bf 2,1})$
but no singlets charged under $U(1)_{B-L}$.\footnote{$^1$}{For all tori except
the ${\bf AAA}$ type, one can even set up 
D6-brane configurations without tachyons at all.} 
Actually, we find a couple of hundred models having either a Higgs
in $({\bf 2,1})$ or a Higgs in the `symmetric' representation of
$U(1)_{B-L}$. But no model contains both Higgs fields. 
This looks discouraging at first sight. Regarding the necessity to have a
hierarchy of a high scale breaking of $U(1)_{B-L}$ and low scale electroweak
Higgs mechanism, we are forced to choose a model with a singlet Higgs
condensing at the
string scale but without Higgs field in the $({\bf 2,1})$ and favour an
alternative mechanism for electroweak symmetry breaking. 
An explicit realization is for example given by 
\eqn\example{ \eqalign{ 
[(n_1^I, m_1^I)] &= [(-3,2),(0,1),(0,-1)] , \cr
[(n_2^I, m_2^I)] &= [(-3,2),(0,1),(0,-1)] , \cr
[(n_3^I, m_3^I)] &= [(-3,2),(1,-1),(-1,0)]. \cr
}}
This model has precisely 3 Higgs singlets 
\eqn\higgs{ h_i : \quad ({\bf 1,1})_{(0,0,-2)} }
which carry only $B-L$ but no hypercharge. 
They are former `superpartners' of the right-handed neutrinos.   
Interestingly, it turns out that all solutions to the tadpole conditions
which display the Higgs singlet charged under $U(1)_{B-L}$ and no tachyons
otherwise result from a model with gauge group $SU(5)\times U(1)$ deformed by
giving a vacuum expectation value to a scalar in the adjoint of
$SU(5)$. Geometrically this is evident in the fact that the stacks of branes
that support the $SU(3)$ and $SU(2)_L$ are always parallel, thus their 
displacement is a marginal deformation at tree level. Of course, we
have to expect that quantum corrections will generate a potential for the
respective adjoint scalar.   

\subsec{An SU(5)$\times$U(1) GUT model}
\noindent
In this section we reinterpret the above 
direct realization of the extended Standard Model
as a GUT scenario. The unified model basically consists in moving the two
stacks for the $U(2)$ and $U(3)$ sector on top of each other, thus tuning the
adjoint Higgs {\bf 24} to a vanishing vacuum expectation value. 
The common GUT gauge group $SU(5)$ is 
extended by a single gauged $U(1)$ symmetry. On two stacks of branes
with $N_{5}=5$ and  $N_{1}=1$ the model is realized by picking again 
\eqn\threegen{ (Y_5,Z_5)=\left(3,{1\over 2}\right), \quad
               (Y_1,Z_1)=\left(3,-{1\over 2}\right) .}
The task of expressing these effective winding numbers in terms of
$[(n_a^I,m_a^I)]$ quantum numbers is identical to that for the previously
discussed extended Standard Model. The number of solutions is again 36 per
stack, i.e. the total set consists of $4\cdot 36^2$ inequivalent models.   
The resulting spectrum of net chiral fermions is featured in table 4.  
\vskip 0.8cm
\vbox{
\centerline{\vbox{
\hbox{\vbox{\offinterlineskip
\def\tablespace{height2pt&\omit&&\omit&&\omit&&
 \omit&\cr}
\def\tablerule{\tablespace\noalign{\hrule}\tablespace}

\hrule\halign{&\vrule#&\strut\hskip0.2cm\hfill #\hfill\hskip0.2cm\cr
\tablespace
& Number && $SU(5) \times U(1)^2$ && $U(1)_{\rm free}$ &\cr
\tablerule
& $3$ && $(\o{\bf 5},{\bf 1})_{(-1,1)}$ &&  $-{6\over 5}$ & \cr
\tablespace
&  $3$ && $({\bf 10},{\bf 1})_{(2,0)}$ && ${2\over 5}$ & \cr
\tablespace
& $3$ && $ ({\bf 1},{\bf 1})_{(0,-2)}$ && ${-2}$ & \cr
}\hrule}}}}
\centerline{
\hbox{{\bf Table 4:}{\it ~~ Left-handed fermions for the 3 generation 
$SU(5) \times U(1)$ model.}}}
}
\vskip 0.5cm 
\noindent
The anomalous $U(1)$ is given by 
\eqn\anomalous{    U(1)_{\rm mass}=5\, U(1)_5 + U(1)_1, }
in accord with \massiv, and the anomaly-free one is 
\eqn\anomalous{ \eqalign{   U(1)_{\rm free} = {1\over 5}\, U(1)_5 - U(1)_1
. \cr  }} 
This is the desired field content of a grand unified Standard Model with extra
right-handed neutrinos, which then also fits into $SO(10)$ representations. 
The usual minimal Higgs sector consists of the adjoint {\bf 24} to break
$SU(5)$ to $SU(3)\times SU(2)_L \times U(1)_Y$ and a $({\bf 5,1})$
which produces the electroweak breaking. In addition we now also need to have
a singlet to break the extra $U(1)_{\rm free}$ gauge factor. 
The adjoint scalar is present as part of the 
vectormultiplet of the formerly ${\cal N}=4$ supersymmetric sector of strings
starting and ending on identical branes within the stack $[5]$. 
Turning on vacuum expectation
values in the supersymmetric theory means moving on 
the Coulomb branch of the moduli space, which 
geometrically translates to separating the 5 D6$_5$-branes into 
parallel stacks of 2 plus 3. Actually, the form of the 
potential generated for this modulus after supersymmetry
breaking is not known, and the existence of a negative mass term as required
for the spontaneous condensation remains speculative. 

Having identified the $SU(5)$ GUT as a Standard Model where two stacks of
branes are pushed upon each other, we can refer to the former analysis of the
scalar spectrum for the other two Higgs fields needed. 
The results of our search for Higgs singlets and bifundamentals done for the $SU(3)
\times SU(2)_L \times U(1)_Y \times U(1)_{B-L}$ model in the previous chapter 
apply without modification as the two stacks for $SU(3)\times SU(2)_L$ 
are parallel in all cases. 

\subsec{Yukawa couplings} 
\noindent 
There is another deviation from the Standard Model which also supports a
replacing of the fundamental Higgs scalar by an alternative composite
operator. Namely,
due to the additional global symmetries, an appropriate Yukawa coupling giving
a mass to the $(u,c,t)$-quarks is absent. 
This can be seen from the quantum numbers of the Higgs field $\tilde H$
resulting from of the relevant Yukawa coupling
\eqn\higgsyu{   \tilde H\, \o{Q}_L\, u_R. }
Thus, $\tilde H$ is forced to have the quantum numbers
\eqn\quantumnum{   \tilde H:\quad ({\bf 1,2})_{(3,1,0)} .}
Apparently, no microscopic open string state can transform in the
singlet representation of $U(3)$ and nevertheless having
$U(1)$ charge $q=3$. 
Note, that for the $(d,s,b)$ quarks and the leptons 
the relevant Yukawa coupling is
\eqn\higgsyub{   H\, \o{Q}_L\, d_R, \quad H\, \o{l}_L\, e_R,
    \quad H^*\, \o{l}_L\, \nu_R   }
leading to the quantum numbers $({\bf 1,2})_{(0,1,1)}$ for the Higgs fields
$H$, which at least is not in contradiction to the open string origin 
of the model. 
We conclude, that in open string models where the $(u,c,t)$-quarks
arise from open strings in the antisymmetric representation
of $U(3)$, there appears a problem with the usual Higgs mechanism.

In a very similar fashion, in the $SU(5)\times U(1)$ GUT model the $U(1)_{\rm mass}$ 
does not allow Yukawa couplings of the type ${\bf 10\cdot 10\cdot 5}$
so that the standard mass generation mechanism does not work. 
Again we are drawn towards a more exotic version of gauge symmetry
breaking and mass generation. 
The only resolution to this obstacle is to propose that the Higgs 
fields are not fundamental fields but composite objects with the
quantum numbers given in \quantumnum.
This possibility is further supported by the analysis in sections 
5.5 and 5.6.

Finally, let us discuss the generation of neutrino masses.
A scalar $h$ in the `symmetric' representation
of the $U(1)_3$ can break the $U(1)_{B-L}$ symmetry via the Higgs mechanism,
but does not
directly lead to a Yukawa coupling of Majorana type for the 
right moving neutrinos. However, the dimension five coupling
\eqn\dimfive{   {1\over M_s}\, \left(h^*\right)^2\, 
            \left(\o\nu^c\right)_L\, \nu_R }
is invariant under all global symmetries and leads to a Majorana
mass for the right moving neutrinos.
Together with the above mentioned (to be found) 
composite Higgs mechanism for the standard Higgs field
this in principle allows the realization of the see-saw mechanism
to generate small neutrino masses. This is in contrast to the
neutrino sector as found in the toroidal models \rimr\ 
where only neutrino masses of Dirac type could be generated due to the
conservation of the lepton number. 

\subsec{Gauge couplings}

\noindent
We now comment on the patterns of gauge coupling unification.  
By dimensional reduction the $U(N_a)$ gauge couplings are given
by
\eqn\gauge{    {4\pi^2\over g_a^2}={M_s\over g_s} \prod_{I=1}^3 L_a^I}
where $g_s$ is the string coupling.
Using for the abelian subgroups $U(1)_a\subset U(N_a)$ the usual 
normalization
tr$(Q_a^2)=1/2$, the gauge coupling for the hypercharge
\eqn\lina{  Q_Y=\sum_a c_a\, Q_a }
is given by 
\eqn\ygauge{   {1\over g_Y^2}=\sum_a {1\over 4} {c_a\over g_a^2} .}
Thus, in our case we get with 
\eqn\linb{  Q_Y=-{2\over 3} U(1)_1 + U(1)_2 =
         -{2\over 3}\sqrt{6}\, \widetilde{U(1)}_1 + 2\,\widetilde{U(1)}_2}
for the Weinberg angle
\eqn\weinberg{  \sin^2 \vartheta_W={3\over 6+2{g_2\over g_1}} .}
The tilded $U(1)$s in \linb\ denote the correctly normalized
ones. 
Since  in all interesting cases  the $U(3)$ branes have the same internal 
volumes than
the $U(2)$ branes, \weinberg\  reduces to
 the prediction $\sin^2 \vartheta_W=3/8$,
which is precisely the $SU(5)$ GUT result. 
Note, that in contrast to the toroidal intersecting brane world
scenario, here the Weinberg angle is completely fixed by the wrapping numbers
of the D6-branes. Thus, these models are more predictive and, 
of course, easier to falsify. 

As usual, in order for the gauge couplings and the string coupling 
to be of order one at the string scale, 
the sizes of the tori are forced to be of order the string scale. 
In principle, by blowing
up the 27 orbifold fixed points we can realize  a large extra dimension
scenario with arbitrary string scale. 
The gauge couplings remain of order one as the D6-branes 
can avoid the blown up $\IP^1$s, whereas the Planck scale gets large 
due to the large overall volume of the compactification manifold.
However, the above result for the Weinberg angle rather suggests that the 
string scale is close to the GUT scale.

In order to compare the gauge couplings
to their experimental values at the weak scale, one has to include in the
beta-function all states with masses between the weak and the string scale.
For a detailed analysis we would need the precise masses of all fields.
Some of these masses are due to loop  correction as for instance
for  the superpartners in the ${\cal N}=4$ vectormultiplet, other
masses are already there at tree level like for the lowest energy scalars
in the non-supersymmetric open string NS sectors. 
 
Thus, the Standard Model gauge couplings run up to
the string scale in the same way as in the non-supersymmetric 
$SU(5)$ GUT model. However, around 1 or 2 orders of magnitude
below the string scale a lot of new states will begin to contribute to the
beta-function and change the running considerably. Since all
the states from the  ${\cal N}=4$ vectormultiplets might  contribute
we expect the one-loop beta-function even to change sign.
Thus, at least in principle it is not excluded that the 
non-supersymmetric $SU(5)$ model will feature gauge coupling
unification.

\subsec{Proton decay}

\noindent
In the three generation models in \rimr\  
the decay of the proton was prohibited, as
the baryon and lepton numbers survived 
the Green-Schwarz mechanism as separate global symmetries. 
In our orbifold models only the combination $B-L$ 
appears as a symmetry, so that there are potential problems
with the stability of the proton. 

In \rafiruph\ it was argued that perturbatively the proton is stable
in intersecting brane models, as effective couplings with three
quarks are forbidden as long as the quark fields appear
in bifundamental representations of the stringy gauge group. 
Apparently, also this argument does not directly apply to our case. 
Indeed the disc diagram in figure 5  generates a dimension six coupling
\eqn\disccoup{ {\cal L}\sim {1\over M_s^2} 
             \left(\o{u}^c_L \, u_L\right)\,
             \left(\o{e}^+_L \,  d_L\right), }
which preserves $B-L$ but violates baryon and lepton numbers separately.
The numbers at the boundary indicate the 
D6$_a$-brane to which the boundary of the disc is attached. 
\fig{}{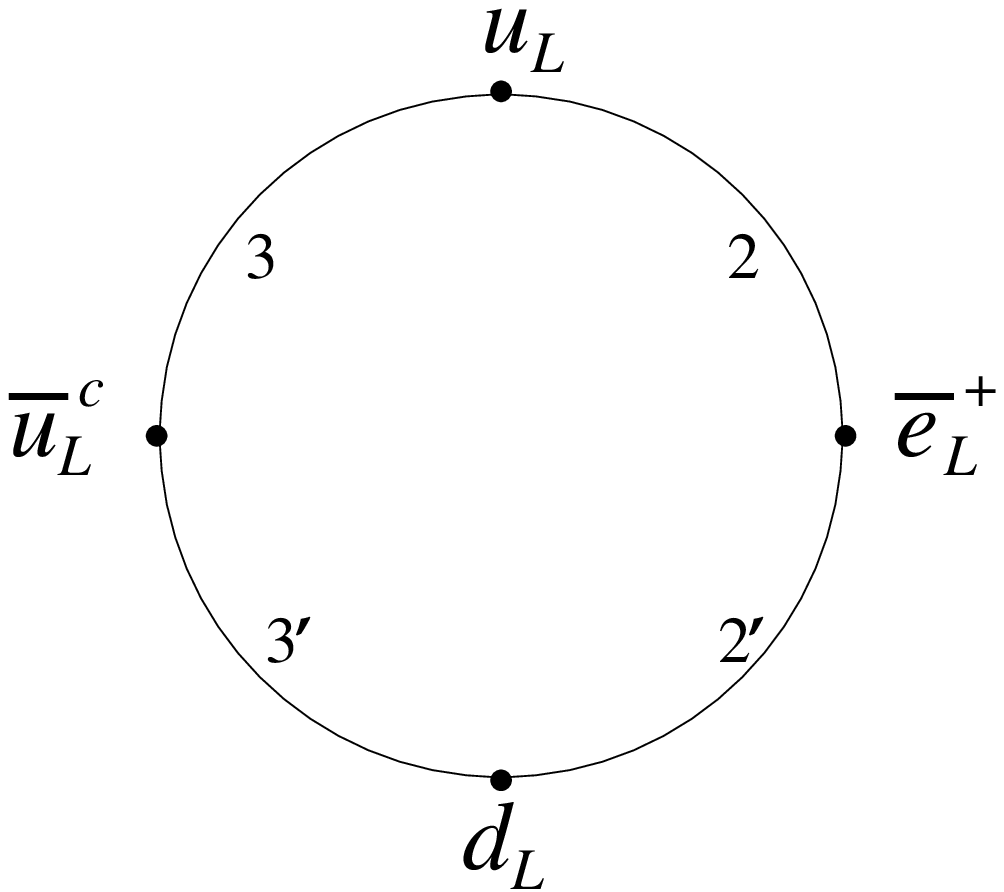}{6truecm}
\noindent
Thus we conclude, that, as long as we want to work in the large extra 
dimension scenario with $M_s\ll 10^{16}$GeV,
these models do have serious problems with proton decay. 
Said differently, also the issue of proton decay leads one 
to chose the string scale rather at the GUT scale than in the 
TeV region.

\newsec{Conclusions}
\noindent
In this paper we have discussed the issue of stability of toroidal
intersecting brane worlds to the
next to leading order in string perturbation theory. 
The arising instability for the geometric parameters
of the internal torus was cured in the case of specific
orbifold models where the complex structure is completely frozen.
We have studied such a partly stabilized 
$\ZZ_3$ orbifold model in great detail and focused
on the derivation of the Standard Model.
It was possible in this, compared to the toroidal case,
fairly constrained set-up to construct a three generation Standard Model
extended by right-handed neutrinos and a gauged $U(1)_{B-L}$ symmetry. 
A detailed study of the tachyon spectra of these models revealed that it was
impossible to realize the entire minimal set of Higgs scalars as open string
tachyons. Thus we had to propose an alternative composite operator instead of
the fundamental Higgs field to achieve the electroweak symmetry
breaking. Furthermore, it turned out that all the models with the required
Higgs to break the $U(1)_{B-L}$ at the string scale are obtained from an
$SU(5)\times U(1)$ GUT model by the condensation of an adjoint Higgs which is
massless at tree level. Thus, the present class of models appears to be 
naturally unified in terms of the common $SU(5)$ scenario. 

The analysis of more detailed phenomenological issues uncovered 
some important deviations from the ordinary Standard Model due to extra global
symmetries, remnants of a larger gauge symmetry after a  
Green-Schwarz anomaly cancellation mechanism. 
The most serious one is surely the absence of appropriate Yukawa couplings 
for the $(u,c,t)$-quarks to  generate  masses via  a fundamental Higgs 
condensation. It appears to be a generic feature that
whenever the Standard Model or GUT matter fields do not
exist in the string spectrum as bifundamental fields exclusively, global
symmetries forbid some of 
the Yukawa couplings required in the standard mass generation
process. Therefore, the only resolution seems to consist of 
a model with composite Higgs which presumably circumvents the gauge hierarchy
problem simultaneously. Unfortunately, there is no direct way to prove the
existence of such a composite operator with a condensate at the electroweak
scale.    

Moreover, without the string scale being
of the same order as the GUT scale there would appear 
problems with  proton-decay
and gauge coupling unification. Thus, it appears that the natural scale
for this intersecting brane model on an orbifold is not the
TeV scale but the GUT scale.

\vskip 1cm

\centerline{{\bf Acknowledgments}}\pano
We would like to thank Lars G\"orlich, 
Axel Krause and Angel Uranga for helpful discussions and correspondence. 
The group is supported in part by the EEC contract ERBFMRXCT96-0045. 
B.~K. and T.~O. also like to thank the Graduiertenkolleg {\it The Standard
Model of Particle Physics - structure, precision tests and extensions},
maintained by the DFG.  
In addition this research was supported in part by the National Science
Foundation under Grant No. PHY99-07949 through the Institute for
Theoretical Physics in Santa Barbara. R.~B. and D.~L. thank the ITP
for the hospitality during this work.

\vfill\eject
\appendix{A}{Effective wrapping numbers}

In this appendix we present the precise form of the effective wrapping
numbers, $Y_{[a]}$ and $Z_{[a]}$, in terms of the fundamental wrapping
numbers, $(n^I_a, m^I_a)$.
\bigno
{\bf AAA} torus
\bigno
\eqn\AAA{ \eqalign{
      Z_{[a]}=&\, n^1_a\, n^2_a\, n^3_a+ {1\over 2} m^1_a\, n^2_a\, n^3_a +
         {1\over 2} n^1_a\, m^2_a\, n^3_a + {1\over 2} n^1_a\, n^2_a\, m^3_a
    -{1\over 2} m^1_a\, m^2_a\, n^3_a  \cr
          & - {1\over 2} m^1_a\, n^2_a\, m^3_a -
    {1\over 2} n^1_a\, m^2_a\, m^3_a - m^1_a\, m^2_a\, m^3_a \cr
     Y_{[a]}=&\, n^1_a\, m^2_a\, m^3_a+m^1_a\, n^2_a\, m^3_a+ m^1_a\, m^2_a\, n^3_a+
         n^1_a\, n^2_a\, m^3_a+ n^1_a\, m^2_a\, n^3_a +m^1_a\, n^2_a\, n^3_a\cr}}
\bigno
{\bf AAB} torus
\bigno
\eqn\AAB{ \eqalign{
      Z_{[a]}=&\, n^1_a\, n^2_a\, n^3_a+ {1\over 2} m^1_a\, n^2_a\, n^3_a +
         {1\over 2} n^1_a\, m^2_a\, n^3_a + {1\over 2} n^1_a\, n^2_a\, m^3_a
    -{1\over 2} m^1_a\, m^2_a\, n^3_a - m^1_a\, m^2_a\, m^3_a \cr
     Y_{[a]}=&\, m^1_a\, m^2_a\, m^3_a+ 2\, n^1_a\, m^2_a\, m^3_a+
        2\, m^1_a\, n^2_a\, m^3_a + n^1_a\, n^2_a\, m^3_a+
         3\,  m^1_a\, m^2_a\, n^3_a \cr
         & + 3\, n^1_a\, m^2_a\, n^3_a + 
       3\,m^1_a\, n^2_a\, n^3_a \cr}}
\bigno
{\bf ABB} torus
\bigno
\eqn\ABB{ \eqalign{
      Z_{[a]}=&\, n^1_a\, n^2_a\, n^3_a+ {1\over 2} n^1_a\, n^2_a\, m^3_a +
         {1\over 2} n^1_a\, m^2_a\, n^3_a + {1\over 2} m^1_a\, n^2_a\, n^3_a
    +{1\over 6} n^1_a\, m^2_a\, m^3_a -{1\over 6} m^1_a\, m^2_a\, m^3_a \cr
     Y_{[a]}=&\, 3\biggl( m^1_a\, m^2_a\, m^3_a+ n^1_a\, m^2_a\, m^3_a+
        2\, m^1_a\, n^2_a\, m^3_a + 2\, m^1_a\, m^2_a\, n^3_a+
         n^1_a\, n^2_a\, m^3_a \cr
     &\phantom{\, 3ttt}   + n^1_a\, m^2_a\, n^3_a + 3\,m^1_a\, n^2_a\,
n^3_a\biggr) \cr}}
\bigno
{\bf BBB} torus
\bigno
\eqn\BBB{ \eqalign{
      Z_{[a]}=&\, n^1_a\, n^2_a\, n^3_a+ {1\over 2} n^1_a\, n^2_a\, m^3_a +
         {1\over 2} n^1_a\, m^2_a\, n^3_a + {1\over 2} m^1_a\, n^2_a\, n^3_a
    +{1\over 6} n^1_a\, m^2_a\, m^3_a \cr
       & + {1\over 6} m^1_a\, n^2_a\, m^3_a +
       {1\over 6} m^1_a\, m^2_a\, n^3_a \cr
     Y_{[a]}=&\, 3\biggl(2\, m^1_a\, m^2_a\, m^3_a+ 3\, n^1_a\, m^2_a\, m^3_a+
        3\, m^1_a\, n^2_a\, m^3_a + 3\, m^1_a\, m^2_a\, n^3_a+
         3\, n^1_a\, n^2_a\, m^3_a \cr
 &\phantom{\, 3ttt} + 3\, n^1_a\, m^2_a\, n^3_a + 3\,m^1_a\, n^2_a\, n^3_a \biggr)\cr}}

\vfill\eject
\appendix{B}{Intersecting branes on the 6D $\ZZ_3$ orbifold}

In this appendix we summarize the results for the tadpole
cancellation conditions and the massless spectra for the
six-dimensional $\ZZ_3$ orbifolds. 
The orbifold action on two complex coordinates is 
\eqn\act{ Z_1\to e^{2\pi i / 3}\, Z_1,\quad\quad 
          Z_2\to e^{-{2\pi i / 3}}\, Z_2 .}
As in the four-dimensional case we can distinguish between the
two differently oriented tori, ${\bf A}$ and ${\bf B}$, so that in this case
we get the three different models, ${\bf AA}$, ${\bf BB}$ and ${\bf AB}$. 
For the six-dimensional closed string spectrum with ${\cal N}=(0,1)$
supersymmetry one gets besides the supergravity multiplet
\eqn\closed{\eqalign{ &{\bf AA}:\ 8\times {\rm tensors}+13\times {\rm hypers}
,\cr 
                      &{\bf AB}:\ 6\times {\rm tensors}+15\times {\rm hypers}
,\cr 
                      &{\bf BB}:\ 21\times {\rm hypers}. \cr}}
Similar to the four-dimensional case we can define the following
quantity
\eqn\yizib{\eqalign{
       &Z_{[a]}={1\over 3} \sum_{({n}^I_b,{m}^I_b)\in [a]}
                           \prod_{I=1}^2  \left( {n}^I_b +{1\over 2}\,
                         {m}^I_b \right) \cr }}
which for the three different tori read
\eqn\zzzz{\eqalign{
   &{\bf AA:}\quad  Z_{[a]}=n^1_a\, n^2_a+{1\over 2} n^1_a\, m^2_a +
               {1\over 2} m^1_a\, n^2_a-{1\over 2} m^1_a\, m^2_a ,\cr
   &{\bf AB:}\quad  Z_{[a]}=n^1_a\, n^2_a+{1\over 2} n^1_a\, m^2_a +
               {1\over 2} m^1_a\, n^2_a+{1\over 6} m^1_a\, m^2_a ,\cr
   &{\bf BB:}\quad  Z_{[a]}=n^1_a\, n^2_a+{1\over 2} n^1_a\, m^2_a +
               {1\over 2} m^1_a\, n^2_a  .}}
Then, the RR-tadpole cancellation condition can be expressed as
\eqn\rrtads{  \sum_a N_a\, Z_{[a]} =4 .}
Moreover, we define
 \eqn\yyyy{\eqalign{
   &{\bf AA,BB:}\quad  Y_{[a]}= n^1_a\, m^2_a +
               m^1_a\, n^2_a+m^1_a\, m^2_a ,  \cr
   &{\bf AB:\phantom{BB}}\quad  Y_{[a]}={1\over 2} n^1_a\, m^2_a +
               {3\over 2} m^1_a\, n^2_a+ m^1_a\, m^2_a }}
and the $L^I_a$ as in equation \lls. 
Then the chiral  massless spectra in the $({\bf 1,2})$ representation
of the little group $SO(4)=SU(2)\times SU(2)$ for the three 
different four-dimensional tori read:
\meno
{\bf AA} torus

\vbox{
\centerline{\vbox{
\hbox{\vbox{\offinterlineskip
\def\tablespace{height2pt&\omit&&
 \omit&\cr}
\def\tablerule{\tablespace\noalign{\hrule}\tablespace}

\hrule\halign{&\vrule#&\strut\hskip0.2cm\hfil#\hfill\hskip0.2cm\cr
\tablespace
& Rep.  && Number  &\cr
\tablerule
& $(N_a,\o N_b)+ c.c. $ &&  $2Z_{[a]}\, Z_{[b]}+
             {3\over 2} Y_{[a]}\, Y_{[b]} $ &  \cr
\tablespace
& $(N_a,N_b)+ c.c. $ &&  $2Z_{[a]}\, Z_{[b]} -{3\over 2} Y_{[a]}\, Y_{[b] }$ &
 \cr
\tablespace
& $ A_a + c.c. $ &&  $2Z_a $ & \cr
\tablespace
& $ A_a + S_a + c.c. $ &&  $2Z_{[a]}^2-Z_{[a]} - \prod_I \left(L^I_a\right)^2$
 & \cr
\tablespace
& $ {\rm Adj}_a+ c.c. $ &&  $ \prod_I \left(L^I_a\right)^2 $ & \cr
}\hrule}}}}
\centerline{
\hbox{{\bf Table B1:}{\it ~~ Chiral fermions for the {\bf AA} torus .}}}
}
\noindent
\bigno
{\bf AB} torus

\vbox{
\centerline{\vbox{
\hbox{\vbox{\offinterlineskip
\def\tablespace{height2pt&\omit&&
 \omit&\cr}
\def\tablerule{\tablespace\noalign{\hrule}\tablespace}

\hrule\halign{&\vrule#&\strut\hskip0.2cm\hfil#\hfill\hskip0.2cm\cr
\tablespace
& Rep.  && Number  &\cr
\tablerule
& $(N_a,\o N_b)+ c.c. $ &&  $6Z_{[a]}\, Z_{[b]} +2  Y_{[a]}\, Y_{[b]} $ & \cr
\tablespace
& $(N_a,N_b)+ c.c. $ &&  $6Z_{[a]}\, Z_{[b]} - 2 Y_{[a]}\, Y_{[b]} $ & \cr
\tablespace
& $ A_a + c.c. $ &&  $6Z_{[a]} $ & \cr
\tablespace
& $ A_a + S_a + c.c. $ &&  $6Z_{[a]}^2-3Z_{[a]} - 3\prod_I \left(L^I_a\right)^2 $ & \cr
\tablespace
& $ {\rm Adj}_a+ c.c. $ &&  $3 \prod_I \left(L^I_a\right)^2 $ & \cr
}\hrule}}}}
\centerline{
\hbox{{\bf Table B2:}{\it ~~ Chiral fermions for the {\bf AB} torus .}}}
}
\noindent
\bigno
{\bf BB} torus

\vbox{
\centerline{\vbox{
\hbox{\vbox{\offinterlineskip
\def\tablespace{height2pt&\omit&&
 \omit&\cr}
\def\tablerule{\tablespace\noalign{\hrule}\tablespace}

\hrule\halign{&\vrule#&\strut\hskip0.2cm\hfil#\hfill\hskip0.2cm\cr
\tablespace
& Rep.  && Number  &\cr
\tablerule
& $(N_a,\o N_b)+ c.c. $ &&  $18Z_{[a]}\, Z_{[b]} +
                     {3\over 2} Y_{[a]}\, Y_{[b]} $ & \cr
\tablespace
& $(N_a,N_b)+ c.c. $ &&  $18Z_{[a]}\, Z_{[b]} - 
 {3\over 2} Y_{[a]}\, Y_{[b]} $ & \cr
\tablespace
& $ A_a + c.c. $ &&  $18Z_{[a]} $ & \cr
\tablespace
& $ A_a + S_a + c.c. $ &&  $18Z_{[a]}^2-9Z_{[a]} - 9\prod_I \left(L^I_a\right)^2 $ & \cr
\tablespace
& $ {\rm Adj}_a+ c.c. $ &&  $ 9\prod_I \left(L^I_a\right)^2 $ & \cr
}\hrule}}}}
\centerline{
\hbox{{\bf Table B3:}{\it ~~ Chiral fermions for the {\bf BB} torus .}}}
}
\vskip 0.5cm \vfill\eject
\noindent
Let us check explicitly the cancellation of the $F^4$ and $R^4$ anomaly to
provide an additional check of the consistency of the construction. 
For the $F^4$ anomaly of the $U(N_a)$ gauge group we get
\eqn\ffour{\eqalign{
   &3^{{\bf n}_b}\left(\sum_{b\ne a} 4\, N_b\, Z_{[a]}\, Z_{[b]} + 
     2 (N_a-8) Z_{[a]} +
           2\, N_a (2\, Z_{[a]}^2-Z_{[a]}) \right) =   \cr  
    &3^{{\bf n}_b}\left(   4\, Z_{[a]} (4-N_a\, Z_{[a]}) - 16\, Z_{[a]} +
            4\, N_a \, Z_{[a]}^2\right)=0. \cr}}
The $R^4$ anomaly reads
\eqn\frfour{\eqalign{
   &3^{{\bf n}_b}\left({1\over 2}\sum_{b\ne a} 4\, N_a\, N_b\, Z_{[a]}\, 
    Z_{[b]} +  \sum_a {N_a(N_a-1)\over 2}\, 2\, Z_{[a]} +
           \sum_a  N_a^2 (2\, Z_{[a]}^2-Z_{[a]}) \right) = \cr
    &3^{{\bf n}_b}\left({1\over 2}\sum_{a} 4\, N_a\, Z_{[a]} (4-N_a\, Z_{[a]})-
     \sum_a  N_a\, Z_{[a]} +  \sum_a  2\, N_a^2\, Z_{[a]}^2 \right)=
     28\cdot 3^{{\bf n}_b}, \cr}}
which is precisely what one needs to cancel the $R^4$ anomaly resulting from 
the closed string spectrum in \closed.

\listrefs

\bye
\end